\documentclass{article}
\usepackage{graphicx} 

\usepackage[accepted]{icml2026}
\usepackage{xurl}
\usepackage[breaklinks]{hyperref}
\usepackage{tabularx}
\usepackage{booktabs}

\usepackage{tcolorbox}
\tcbuselibrary{many}
\usepackage{enumitem}

\definecolor{myGray}{HTML}{222222}

\tcbset{
  summarybox/.style={
    breakable,
    width=\linewidth,
    top=8pt,
    bottom=4pt,
    colback=green!6!white,
    colframe=myGray,
    colbacktitle=myGray,
    enhanced,
    center,
    attach boxed title to top left={yshift=-0.1in,xshift=5pt},
    boxed title style={boxrule=0pt,colframe=white,
    left=2pt,
    right=2pt,
    top=1pt,
    bottom=2pt},
    left=5pt,
    right=6pt,
    before upper={\setlength{\parskip}{\medskipamount}},
    },
  takeawaybox/.style={
    breakable,
    pad at break*=3pt,
    width=\linewidth,
    top=8pt,
    bottom=4pt,
    colback=blue!6!white,
    colframe=myGray,
    colbacktitle=myGray,
    enhanced,
    center,
    attach boxed title to top left={yshift=-0.1in,xshift=5pt},
    boxed title style={boxrule=0pt,colframe=white,
    left=2pt,
    right=2pt,
    top=1pt,
    bottom=2pt},
    left=5pt,
    right=6pt,
    before upper={\setlength{\parskip}{\medskipamount}},
    before skip=1em,
    after  skip=1em,
    },
  problembox/.style={
    width=\linewidth,
    top=8pt,
    bottom=4pt,
    colback=black!6!white,
    colframe=myGray,
    colbacktitle=myGray,
    enhanced,
    center,
    attach boxed title to top left={yshift=-0.1in,xshift=5pt},
    boxed title style={boxrule=0pt,colframe=white,
    left=0pt,
    right=2pt,
    top=-1pt,
    bottom=-2pt},
    left=5pt,
    right=6pt,
    before skip=1em,
    after  skip=1em,
  },
  positionbox/.style={
    width=\linewidth,
    top=8pt,
    bottom=4pt,
    colback=white!6!white,
    colframe=myGray,
    colbacktitle=myGray,
    enhanced,
    center,
    attach boxed title to top left={yshift=-0.1in,xshift=0.15in},
    boxed title style={boxrule=0pt,colframe=white,},
    left=8pt,
    right=8pt
  },
}
\newtcolorbox{posbox}[2][]{positionbox,title=#2,#1}
\newtcolorbox{mybox}[2][]{takeawaybox,title=#2,#1}
\newtcolorbox{myboxbroken}[3][]{takeawaybox,title=#2,title after break=#3,#1}
\newtcolorbox{pbox}[2][]{problembox,title=#2,#1}
\newtcolorbox{sumbox}[2][]{summarybox,title=#2,#1}

\usepackage{pifont}
\usepackage[dvipsnames]{xcolor}
\newcommand{\cmark}{{\color{ForestGreen}\ding{51}}}
\newcommand{\xmark}{{\color{BrickRed}\ding{55}}}%

\definecolor{colDataBg}{HTML}{0D47A1} 
\definecolor{colDataTx}{HTML}{E3F2FD} 
\definecolor{colEvalBg}{HTML}{E65100} 
\definecolor{colEvalTx}{HTML}{FFF3E0} 
\definecolor{colAdvBg}{HTML}{B71C1C} 
\definecolor{colAdvTx}{HTML}{FFEBEE} 
\definecolor{colWellBg}{HTML}{004D40} 

\newtcbox{\challengechip}[1][]{%
  on line,
  colback=gray!10,
  colframe=gray!40!black,
  boxrule=0.3pt,
  arc=1pt,
  outer arc=1pt,
  boxsep=0.5pt,
  left=3pt,
  right=3pt,
  top=1pt,
  bottom=1pt,
  nobeforeafter,
  tcbox raise base,
  enhanced,
  fontupper=\small\bfseries,
  #1
}

\newcommand{\chipData}{%
  \challengechip[colback=colDataBg, colframe=colDataBg, colupper=white]{DATA}}

\newcommand{\chipEval}{%
  \challengechip[colback=colEvalBg, colframe=colEvalBg, colupper=white]{EVAL}}

\newcommand{\chipAdv}{%
  \challengechip[colback=colAdvBg, colframe=colAdvBg, colupper=white]{ADV}}

\newcommand{\chipWell}{%
  \challengechip[colback=colWellBg, colframe=colWellBg, colupper=white]{WELL}}
  
\newtcolorbox{pboxchipped}[3][]{%
  problembox,%
  title={%
    \if\relax\detokenize{#3}\relax
      #2%
    \else
      \begin{tabular}{@{}l@{}}
        #2\\[0em]
        #3
      \end{tabular}%
    \fi
  },
  #1,
}

\icmltitlerunning{Position: Preventing AIG-CSAM Necessitates New Approaches to AI Safety}

\begin{document}

\twocolumn[
\icmltitle{Position: Preventing AI-Generated CSAM\\Necessitates New Approaches to AI Safety}
\icmlsetsymbol{equal}{*}

\vspace{-0.1in}

\begin{icmlauthorlist}
\icmlauthor{Neil Kale}{equal,CMU}
\icmlauthor{Rebecca Portnoff}{equal,Thorn}
\icmlauthor{Pratiksha Thaker}{CMU}
\icmlauthor{Michael Simpson}{Thorn}
\icmlauthor{Robertson Wang}{Thorn}\\
\icmlauthor{Kevin Kuo}{CMU}
\icmlauthor{Chhavi Yadav}{CMU}
\icmlauthor{Virginia Smith}{CMU}
\end{icmlauthorlist}

\icmlaffiliation{CMU}{Carnegie Mellon University}
\icmlaffiliation{Thorn}{Thorn}
\icmlcorrespondingauthor{Neil Kale}{nkale@cs.cmu.edu}
\icmlcorrespondingauthor{Rebecca Portnoff}{rebecca@wearethorn.org}

\vspace{0.2in}
]

\printAffiliationsAndNotice{The first two authors contributed equally and may list their names interchangeably.}

\begin{abstract}
Modern artificial intelligence (AI) systems  present profound new risks to child safety. AI is increasingly being misused to create AI-generated child sexual abuse material, facilitate child sexual exploitation, and reduce barriers to harm. In this paper, we argue that protecting children from AI-facilitated sexual abuse requires new approaches to AI safety. Existing safety techniques assume data accessibility, transparency, and evaluation practices that are incompatible with the ethical and legal constraints surrounding child sexual abuse material. We examine how these constraints create new technical challenges, such as limitations on dataset auditing, red teaming, and fine-tuning prevention. In turn, we outline \textit{15 open problems} in online child sexual exploitation and abuse across the AI development lifecycle, from dataset curation and model design to deployment and long-term maintenance. We propose targeted recommendations for researchers, developers, and policymakers to bridge the gap between theoretical AI safety and the realities of child protection. Our work aims to reframe preventing AI-facilitated child sexual abuse as a central, safety-critical dimension for AI research, motivating work that translates responsible AI principles into concrete safeguards against the exploitation of children.
\end{abstract}

\vspace{-0.3in}
\section{Introduction}
Artificial intelligence (AI) systems have achieved remarkable capabilities in content creation. 
However, these same capabilities increasingly pose risks to children, including facilitating child sexual abuse and exploitation (CSAE), particularly via the development of AI-generated child sexual abuse material (AIG-CSAM). 
This misuse represents a critical challenge at the intersection of AI safety, child protection, and technical system design---one that most existing AI safety frameworks inadequately address \cite{internet2023ai,iwf2024,paltieli,thiel2023generative}.

The scope and severity of this problem are substantial. The National Center for Missing and Exploited Children (NCMEC) received more than 440,000 reports of AI-generated material related to CSAE in the first half of 2025 alone~\cite{2025report}, compared to 7,000  reports in 2023-2024, and the Internet Watch Foundation (IWF) observed a 400\% increase in AIG-CSAM reports since 2024~\cite{iwfblog,iwf2024}. 
A recent study found that 13\% of U.S. sexual extortion victims reported that the perpetrator used AI to create blackmail material \cite{thorn2025sextortion}, and a 2025 survey of Australian young people found that 41\% of sexual extortion victims reported that the material used to blackmail them had been digitally manipulated~\cite{wolbers2025}. 
In the U.S., 1 in 17 teenagers aged 13--17 report having been victimized by deepfake nude images~\cite{thorn24}. 
AIG-CSAM has thus been recognized by the AI safety community as a critical area for intervention, with immediate, quantifiable marginal risk~\cite{kapoor2024societal}.

However, despite the known risks of AIG-CSAM, many existing techniques from AI safety and secure \& privacy-preserving machine learning cannot be directly used for AIG-CSAM prevention, detection, and mitigation. 
Paradoxically, in part because child safety is an important and established area of concern, there are unique constraints and restrictions surrounding CSAM access/usage.
These restrictions are \emph{ethical and necessary}, but create complications for researchers and practitioners seeking to develop and assess AI safety solutions. This motivates our main position (below), which we explore in detail in this work:


\vspace{-.5mm}
\begin{posbox}{Main Position}{
\textit{Existing AI safety research makes assumptions that do not match the legal and ethical constraints of safety applications preventing child sexual abuse and exploitation. Bridging this gap by solving open sociotechnical problems will be critical to make AI safety solutions effective for child safety in practice.}}
\end{posbox}
\vspace{-.5mm}

\begin{table*}[t!]
    \centering
    \begin{tabular}{r  cccc}
        \toprule
          & \textbf{General Public} & \textbf{AI Developers/Providers} & \textbf{Reporting Hotlines} & \textbf{Law Enforcement} \\
        \midrule
        Can legally access CSAM & No & No* & Yes & Yes \\
        Can access hashes & No & Yes & Yes & Yes \\
        Can train on CSAM & No & No* & Yes & Yes \\
        Can generate CSAM & No & No & No & Yes \\
         \bottomrule
    \end{tabular}
    \caption{CSAM access, usage, and generation capabilities for various actors. The inability to access CSAM as well as generate and train on CSAM data can limit the application of existing tools in AI safety. *While AI developers and providers generally can't access or train on CSAM, there are specific, limited cases where access may be permissible (see Section~\ref{background:access}).}
    \label{tab:csam_access}
    \vspace{-.2in}
\end{table*}

We outline current industry consensus on solutions to address AIG-CSAM~\cite{thornsafetybydesign}, pointing out gaps between {assumptions} made by state-of-the-art mitigations 
and {practical} limitations. We then highlight \textit{15 open problems} (9 in the body, 6 in Appendix \ref{app:additional_openproblems}) spanning model development ($\S$\ref{sec:develop}), deployment ($\S$\ref{sec:deploy}), and maintenance ($\S$\ref{sec:maintain}), which represent critical research and policy priorities where advances could have immediate impact on protecting children. Finally, we provide recommendations for researchers, AI providers, and policymakers to address these open problems and bridge ecosystem gaps.

\section{Background}
Generative AI enables non-experts to create realistic digital media at unprecedented scale. The AI safety community has raised concerns about resulting risks, such as bias propagation~\cite{birhane2023into}, disinformation~\cite{musser2023cost}, bioterrorism~\cite{peppin2025reality}, job displacement~\cite{hazra2025position}, and cybercrime~\cite{hazell2023spear}. 

In this work, we specifically examine generative AI risks for \textit{child safety}, focusing on photo-realistic child sexual abuse material (CSAM). AIG-CSAM presents critical harms to child safety. It complicates victim identification by expanding the volume of material that law enforcement must navigate to locate children in abuse scenarios~\cite{thiel2023generative,internet2023ai}. It enables re-victimization as bad actors can fine-tune models on existing CSAM to generate content imitating specific victims while producing new poses and acts of violence~\cite{thiel2023generative,internet2023ai}. It also broadens the pool of victims for sexual extortion, with offenders using image editing tools to sexualize benign depictions of children for such schemes~\cite{fbiPSA}. 


Both preventive and reactive measures around AIG-CSAM are necessary to improve child safety: civil society~\cite{thornsafetybydesignblog}, industry~\cite{meta}, and regulatory bodies~\cite{TakeItDownAct} have all pursued relevant efforts. However, as we show, CSAM imposes unique constraints on existing AI safety techniques.  To illustrate the resulting open research gaps, in this section  we first describe what makes AI child safety particularly challenging ($\S$\ref{background:challenges}) and detail key organizations involved in intervention efforts ($\S$\ref{background:access}). 

\textbf{Disclaimer:} Throughout this work, we note that we focus on AI-generated imagery as the key modality of interest, and primarily take a U.S./western perspective when considering legal constraints that may impact safety techniques. We discuss limitations, broader perspectives, and alternate views in $\S$~\ref{sec:alternate} and $\S$~\ref{sec:conclusion}, and provide a detailed discussion of related work in App.~\ref{sec:relatedwork}.

\subsection{What makes AI child safety challenging?}
\label{background:challenges}
The illegal and sensitive nature of CSAM, alongside other guardrails related to children's data, creates unique challenges for AI safety. Below we highlight several underlying challenges which we will revisit throughout the paper:

\chipData\ --- \textbf{Data restrictions.} CSAM is illegal in most countries, with 
legislation such as COPPA and GDPR establishing strict privacy safeguards for children's data~\cite{FTC_COPPA,gdpreu2018whatgdpr,USC18_possession}. As we show, these data access restrictions are at odds with many standard AI safety approaches, which require access to training and/or evaluation data.

\chipEval\ --- \textbf{Evaluation restrictions.} Evaluation and red teaming methods typically rely on model prompting to assess model behavior and capabilities, but intentionally generating AIG-CSAM is also illegal in the U.S. ~\cite{USC18_protect}. Similarly, standardized CSAM evaluation datasets do not exist, as it is illegal to access this data. 

\chipAdv\ --- \textbf{Adversarial and opaque environment with strict guarantees.} 
Child safety demands strong guarantees, as AIG-CSAM output is intrinsically illegal in most countries and can directly harm specific children. However, these guarantees can be difficult to achieve, as CSAE offenders collaborate to actively circumvent safeguards, while AI system developers typically only disclose high-level descriptions of safety mechanisms ($\S$\ref{app:deployment}).

\chipWell\ --- \textbf{Wellness implications.} CSAM exposure and the psychological demands of adversarial assessment pose substantial risks, including trauma and PTSD~\cite{spence23}. This can make it challenging to apply safety techniques requiring significant human involvement. 

\vspace{-.05in}
\subsection{Organizations and access assumptions}
\label{background:access}
\vspace{-.05in}
Three key groups make up the AIG-CSAM prevention and response ecosystem (see Table \ref{tab:csam_access} for a summary). 


\textbf{AI Developers and Providers.} AI developers (individuals/organizations that build AI technology) and AI providers (platforms that host  AI tools) both have CSAM reporting and retention obligations under U.S. law~\cite{reportact}. They can access CSAM hashes via institutions like NCMEC for matching and reporting. In certain cases, they may be allowed to extract embeddings from CSAM detected on their platforms to train detection methods~\cite{googlecst}*, but direct access would require partnership with law enforcement or reporting hotlines.

\textbf{Reporting Hotlines.} Hotlines (e.g. NCMEC, IWF) are authorized to receive and process CSAM reports, to facilitate removal from online platforms and support LE investigation and prosecution efforts. These organizations legally house CSAM, enabling hash sharing, victim ID, and abuse prevention efforts. While they can provide scoped, secure access to CSAM to other institutions building CSAM prevention and detection technology, they cannot directly assess generative AI models for CSAM capabilities.

\textbf{Law Enforcement (LE).} Finally, LE has the legal authority to investigate, collect, and analyze CSAM as evidence while working to identify victims, apprehend perpetrators, and disrupt child exploitation networks. However, although they have the ability to train detection methods and assess AI models for CSAM capabilities, they may lack the budget or technical expertise for these efforts.

As we will see, the restrictions surrounding CSAM access for these various groups (particularly for AI developers/providers) have wide-ranging impacts on the use of existing safety tools related to AI system development ($\S$\ref{sec:develop}), deployment ($\S$\ref{sec:deploy}), and maintenance ($\S$\ref{sec:maintain}).

\vspace{-2mm}
\section{Developing Safe Models by Design}
\label{sec:develop}


{Developing} safe generative AI models requires proactively addressing child safety risks before and during training. Large-scale datasets used to train generative models are often scraped from the Internet with \textit{minimal cleaning} \cite{bommasani2021opportunities}, leading to cases where datasets 
used to train popular models 
were found to contain CSAM \cite{magid2024you, thiel2023identifying}.  Generative models may also produce harmful outputs through \textit{concept fusion}, combining attributes from separate training examples (e.g., adult pornographic content and benign child depictions) \cite{okawa2023compositional, zhang2021can}. 
Furthermore, by using generative models to partially edit CSAM, offenders can make it difficult to determine whether material depicts active abuse, recovered victim-survivors, or deepfakes. 


\vspace{-2mm}
\subsection {Current Approaches}
Currently, model developers can use \textit{allowed/disallowed website lists} for data curation to avoid sites hosting CSAM, \cite{thorn2025safetybydesignupdate}, 
and can detect CSAM in training data by hash matching against third-party CSAM databases and using CSAM classifiers \cite{lee2020detecting, googlecst, thornsafetybydesignblog}. One approach to address concept fusion is \textit{NSFW filtering}, which considers removing adult material from datasets prior to model development \cite{thornsafetybydesign}.

Developers also use \textit{red teaming} to find harmful content queries, patching exploits before model release \cite{ganguli2022red, google2023red}. Directly prompting for AIG-CSAM is illegal in the U.S., so developers report either testing \textit{proxy concepts} \cite{thornsafetybydesign} or \textit{avoiding} such testing entirely \cite{grossman2025csam}.

\textit{Content provenance} solutions have also been implemented to support victim ID efforts. These tools  provide a feedback mechanism for researchers, AI providers, and policy makers. By allowing stakeholders to become aware of the models that are producing problematic output, systemic issues can be addressed in existing mitigations. Current approaches include \textit{C2PA} \cite{C2PA_Org} (an open metadata-based standard for digital content origin and edits) and \textit{watermarking} \cite{yu2021artificial, wen2023tree, fernandez2023stable}.



\vspace{-2mm}
\subsection{Open Problems}

\begin{pboxchipped}
{\textbf{Open Problem A1: Partial data cleaning}}
{\chipData\ \chipEval}
{How does data cleaning affect a generative model's ability to depict a concept? To what degree can partial cleaning guarantee that a model cannot depict CSAM?}
\end{pboxchipped}
\label{sec:develop:partial-data-cleaning}

Guaranteeing complete CSAM removal is difficult due to imperfect detection technology, scale of data, and moderator wellness concerns. Determining whether imperfect removal (e.g., 99.9\%) can prevent text-to-image models from learning to generate CSAM is an important open question. However, direct analysis of CSAM filtering strategies is challenging due to data access restrictions.


\emph{Existing work.} LLM \citep{obrien2025deep,maini2025safety} and diffusion studies \cite{nichol2021glide} show that filtering training data can minimize harmful capabilities in the  model, and research in text-to-image generation shows a critical number of training samples are required for concept composition \cite{okawa2023compositional,cretu2025evaluating}.

\emph{Limitations.} AIG-CSAM generation requires strong safety guarantees. While existing work is a starting point to study the effectiveness of data cleaning, the AIG-CSAM problem would benefit from  formal guarantees on a model's ability to generate harmful content. Moreover, because it is illegal for researchers to generate CSAM, it is not yet clear how to exhaustively test models' capabilities to ensure they are safe in these scenarios.  


\begin{pboxchipped}
{\textbf{Open Problem A2: Preventing concept fusion}}
{\chipData\ \chipEval\ \chipAdv}
{Can generative models be selectively blocked from combining high-risk concepts, such as children and NSFW material?}
\end{pboxchipped}
\label{sec:develop:concept-fusion}

A unique concern for CSAM is that harmful content can also be created by \textit{composing} two potentially benign concepts (e.g., benign child depictions and adult content) (see example experiments in $\S$\ref{sec:conceptfusion}). Addressing this via comprehensive data filtering is challenging for similar reasons to CSAM removal. In addition understanding whether partial data cleaning prevents concept fusion,  novel architectures and training paradigms could also be developed to selectively prevent unwanted fusion.

\emph{Existing work.} Prior work includes retrieval-based architectures for revocable sensitive data access \cite{min2023silo} and 
 classifiers that self-identify concepts to generate each output \cite{wang2024disentangled, espinosa2022concept}. 
Recent work also explores concept-based explainability for text-to-image models via sparse autoencoders \cite{tinaz2026emergence, surkov2024one}.

\emph{Limitations.} Most concept-based explainability work requires accessing and generating images to identify interpretable features. Mechanisms relying on isolated sensitive datastores face challenges in reliably classifying sensitive imagery (e.g. NSFW content), may make sacrifices in quality on benign concepts, and are further complicated by entangled concepts (e.g. children's medical imagery). 

\subsubsection{Proof-of-Concept: Concept Fusion}
\label{sec:conceptfusion}
\vspace{-.05in}
A concern we raise in Open Problem 2 is that models may be able to combine high-risk concepts such as children and NSFW material. As shown in Figure 1, we find this to be the case with proxy concepts from CelebA \cite{liu2015deep}: we train diffusion models on separate images of people with blonde hair and eyeglasses, and find that they can generate the combined concept of blondes wearing eyeglasses with \emph{zero} prior examples. Concurrent work from \citet{cretu2025evaluating} similarly uses eyeglasses as a proxy for nudity, and finds that text-to-image models can generate children wearing eyeglasses even with 94\% of child images removed. Developing alternative techniques to prevent concept fusion is thus an important area of future work. See App.~\ref{sec:experiment-details} for full experiment details.
\vspace{.15in}

\vspace{-.1in}
\begin{pboxchipped}
{\textbf{Open Problem A3: Resilience to harmful fine-tuning}}
{\chipData\ \chipEval\ \chipAdv}
{How can generative models be post-trained to prevent fine-tuning on CSAM or simultaneously fine-tuning on multiple CSAM-related concepts?}
\end{pboxchipped}
\label{sec:develop:self-destruction}

Unfortunately, even if base models appear `safe', users can also unlock harmful capabilities through post-training procedures such as fine-tuning. Open-weight models allow users to adapt models on various tasks and are vulnerable to tampering \cite{casper2025open}. CSAM perpetrators use GUI-based LoRA fine-tuning software like ComfyUI \cite{comfy2025} or Ostris \cite{ostris2024aitoolkit} to locally fine-tune open source models on CSAM \cite{Thorn_PAI_CaseStudy_2024}. ``Nudifying'' apps generating illegal deepfake sexual material of children similarly exploit the open source ecosystem, optimizing models for clothing removal and face-swapping \cite{ding2025malicious}. Ideally, text-to-image models would allow benign fine-tuning while preventing these harmful uses. 

\emph{Existing work.} \citet{henderson2023self} propose self-destructing classifiers that degrade when fine-tuned on specific tasks. Similar approaches have been explored for diffusion models \cite{gao2024meta, pan2024leveraging}. 

\emph{Limitations.} Solutions for self-destructing models that obstruct joint fine-tuning on separate concepts (e.g., adult sexual content \textit{and} children) without blocking individual concept fine-tuning are largely unexplored. Similarly, strategies to prevent harmful LoRA fine-tuning (vs. full fine-tuning) are lacking. Methods that build resilience to harmful fine-tuning often require training on obstructed tasks; CSAM data access restrictions make this challenging. 

\textit{We explore further problems in development, such as watermarking and minimizing human exposure, in App.~\ref{app:development}.}

\vspace{-0.5em}
\begin{figure}[t!]
    \centering
    \includegraphics[width=0.45\linewidth]{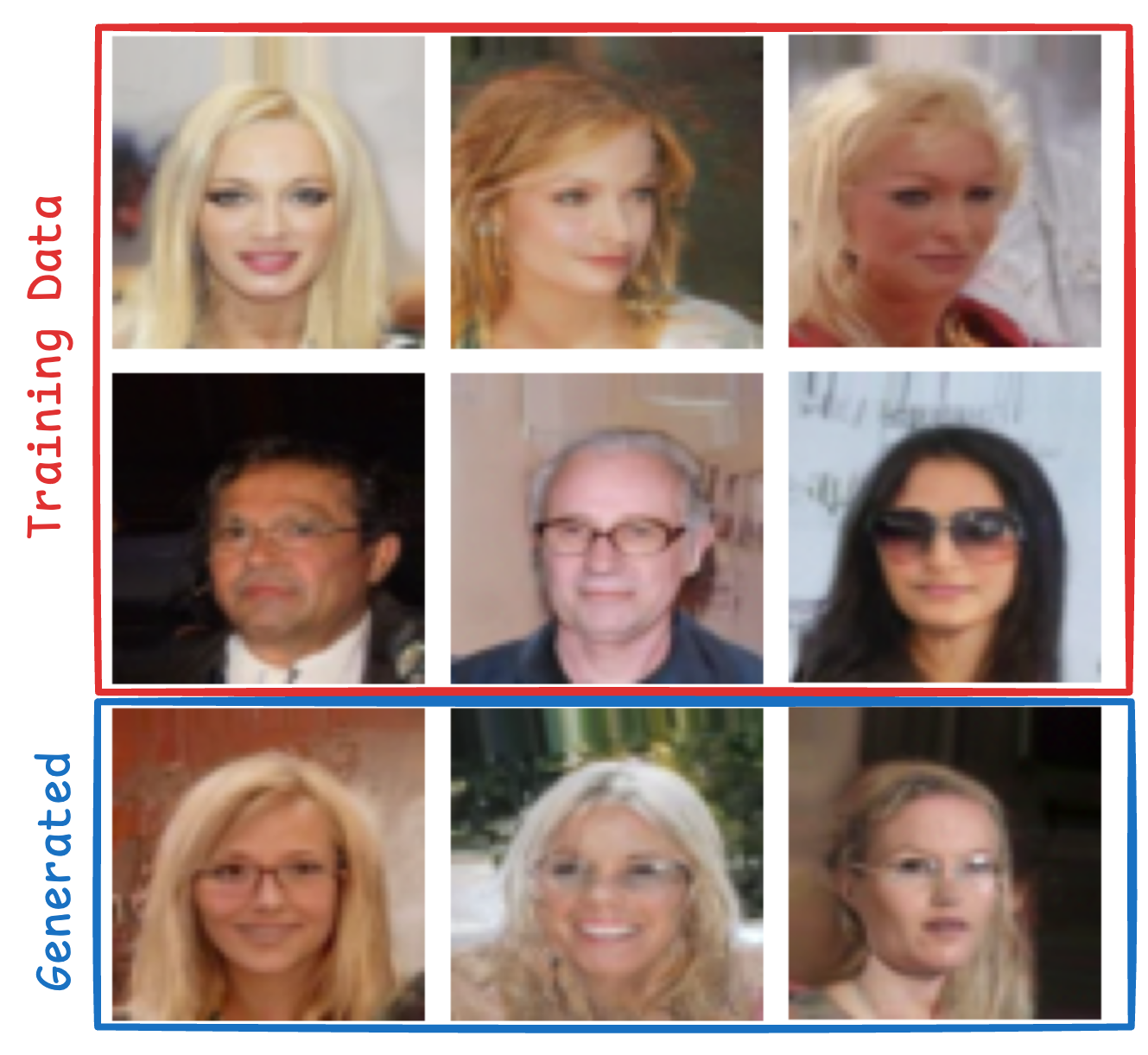}
    \includegraphics[width=0.54\linewidth]{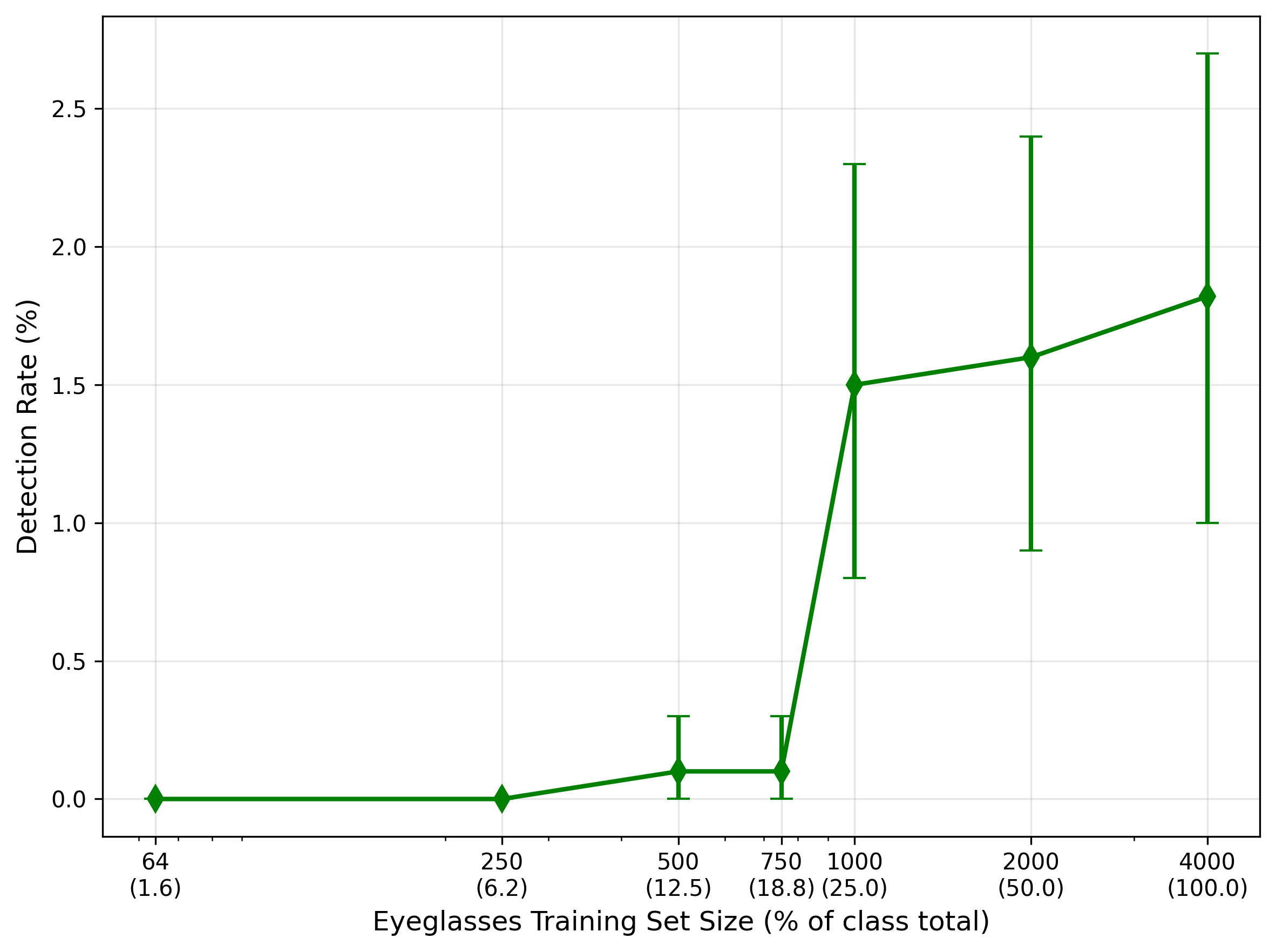}
    \caption{Conditional diffusion models trained on images of blondes (top) and people in eyeglasses (middle) can generate blondes wearing eyeglasses (bottom), \textit{without} overlapping training data. Fusion is effective with a critical threshold of training data (right).}
    \label{fig:composition_samples}
    \vspace{-.2in}
\end{figure}

\subsection{Call to Action}
\vspace{-.05in}
Below we outline take-aways for researchers, AI providers, and policymakers based on the identified open problems. 
\begin{mybox}
{\textbf{Research Directions}}{
\begin{itemize}[wide, labelwidth=0pt, labelindent=0pt, topsep=0pt, partopsep=0pt, parsep=0pt, itemsep=0ex]
    \item Assess partial data cleaning limits using proxy datasets; explore architectures preventing unwanted concept fusion for images/video.
    \item Build self-destructing models targeting the ``nudifying'' ecosystem. 
    \item Create mechanisms to obstruct harmful fine-tuning that do not require access to the harmful material. 
    \item Develop robust content provenance: e.g. models that sharply degrade if finetuned to remove a watermark and localized provenance to handle inpainting.
\end{itemize}
}
\end{mybox}

Given data access restrictions, researchers and AI providers should also establish collaborations with reporting hotlines and LE to enable  effective interventions and evaluation.


\begin{mybox}
{\textbf{Concrete Steps for AI Providers}}{
\begin{itemize}[wide, labelwidth=0pt, labelindent=0pt, topsep=0pt, partopsep=0pt, parsep=0pt, itemsep=0ex]
\item Engage CSAM survivors to understand their perspectives on partial data cleaning and re-victimization. 
\item Partner with hotlines and LE for secure, scoped CSAM access to implement fine-tuning resilience.
\item Reliably label model outputs with C2PA or an equivalent standard and indelible watermarks. For open source models, prioritize tamper-proof content provenance solutions. 
\end{itemize}
}
\end{mybox}

Policymakers play a critical role by establishing long-term pathways to encourage safer model designs. Regulatory efforts, research funding, and private-public partnerships that prioritize child safety are all important for effective action.

\vspace{-.05in}
\begin{mybox}[after skip=0pt]
{\textbf{Policy Goals}}{
\begin{itemize}[wide, labelwidth=0pt, labelindent=0pt, topsep=0pt, partopsep=0pt, parsep=0pt, itemsep=0ex]
\item Establish avenues for scoped, secure evaluation of AIG-CSAM capabilities by vetted institutions. 
\item Resource standards organizations (e.g. NIST) to ensure that their guidance for content provenance in the open source setting remains up to date. 
\item Instruct regulators to engage with the fine-tuning software ecosystem; assess what interventions are in place to prevent the misuse of these tools.
\item Task grant-making institutions to prioritize related research efforts in AI child safety.
\end{itemize}
}
\end{mybox}

\vspace{.1in}
\section{Deployment Safeguards}
\label{sec:deploy}
Once AI models have been trained and evaluated for child safety, they should also be deployed with safeguards  built into the release and distribution processes. Safeguards may include techniques such as content moderation, usage violation detection, reporting mechanisms, and transparency disclosures. These defenses are complicated by the AI model ecosystem, with providers building iteratively off each other's systems via post-training and distillation.


\vspace{-.05in}
\subsection {Current Approaches}
\vspace{-.05in}
Closed-source AI providers currently use \textit{safety classifiers} \cite{inan2023llama} to detect harmful inputs/outputs. They can direct users who attempt to generate illegal content to help resources \cite{thorn2025safetybydesignupdate}. Some open- and closed-source providers establish \textit{user reporting pathways} for violating outputs, prompts, and models. Open-source models may also include safety classifiers \cite{CompVis2022StableDiffusion}, or be \textit{safety fine-tuned} to reject adversarial prompts \cite{kim2024race, gandikota2023erasing}.

To anticipate adversaries who may attempt to circumvent safeguards, providers employ \textit{red teaming}  to identify harmful prompts for training classifiers \cite{ganguli2022red}. However, this is less common on third-party platforms \cite{thorn2025safetybydesignupdate}. While platforms like HuggingFace generally prohibit third-party models with AIG-CSAM capabilities, enforcement of these policies varies \cite{HarrisWillner_StableDiffusionCSAM_2024}. One measure for developer transparency and accountability is recommending \textit{model cards}.

\subsection{Open Problems}
\vspace{-.05in}
Adversaries can jailbreak systems by disguising harmful signals \cite{ma2024jailbreaking, zou2023universal}; text-to-image models are particularly vulnerable \cite{rando2022red,robey2023smoothllm, wei2023jailbreak, jain2023baseline}.

\begin{pboxchipped}
{\textbf{Open Problem B1: Effective prompt/output detection}}
{\chipData\ \chipEval\ \chipAdv \chipWell}
{How can harmful prompts/outputs be reliably detected, in an adversarial ecosystem with limited access to in-distribution examples?}
\end{pboxchipped}

 \vspace{-.05in}
 Open source deployments face additional challenges: content moderation is not feasible, safety filters are vulnerable \cite{zhuang2023pilot}, and fine-tuning interventions are easily reversed~\cite{qifine,huunlearning}.

\emph{Existing work.} Existing harmful prompt refusal requires instruction fine-tuning on similar prompts or few-shot examples in context \cite{wei2023jailbreak, markov2023holistic}. Other solutions include automated red teaming tools to search for violatory prompts \cite{li2024art}, zero-shot prompt classifiers \cite{inan2023llama} and guideline based model training to reject unsafe prompts \cite{guan2024deliberative}. 


\emph{Limitations.} Red teaming prompts and guidelines may not reflect actual CSAM offender behavior. 
Most organizations do not have access to CSAM; accurate classifiers require offender data. For those that do, there is no public benchmark to evaluate their solutions. Research to strengthen solutions in the open source setting is broadly lacking. 



\begin{pboxchipped}
{\textbf{Open Problem B2: Automated model assessment}}
{\chipEval\ \chipAdv \chipWell}
{How can models be assessed for AIG-CSAM capabilities and CSAM training data automatically?}
\end{pboxchipped}


Third-party models constitute a significant chunk of the model ecosystem, and are not always assessed for CSAM risks pre-deployment. Scalable assessment of deployed models is needed to enforce platform policies and prevent distribution of models with AIG-CSAM capabilities.

\emph{Existing work.} Training data extraction techniques \cite{carlini2023extracting} have been applied to detect the presence of specific media in training data. Data attribution techniques for generative models could also prove useful to identify the samples that enable AIG-CSAM \cite{georgiev2023journey, lin2025diffusion, zheng2024intriguing}.
The nascent area of mechanistic interpretability (MI) aims to examine learned weights to understand large models \cite{templeton2024scaling}, with techniques such as automated circuit discovery attempting to identify computational subgraphs that implement specific model behaviors \cite{conmy2023towards}, allowing auditing of specific capabilities by searching relevant subgraphs. 

\emph{Limitations.} Research using MI to audit  CSAM generation capabilities is unexplored; seeking out CSAM trained models to assess such techniques may open  researchers to risk~\cite{thaker2025membership}. 
Current techniques for training data extraction rely on prompting for AIG-CSAM, which violates US law. Text-to-video assessment is also broadly unexplored.

\textit{We explore further problems in deployment, such as model transparency and standardize assessments, in App.~\ref{app:deployment}.}

\subsection{Call to Action}
As with the previous open problems, in the absence of collaborations that enable researchers access to CSAM or CSAM trained models, there are still lines of research that can be pursued that directly ladder up, e.g.:
\begin{mybox}
{\textbf{Research Directions}}
{
\begin{itemize}[itemsep=0ex,leftmargin=*]
    \item Design image-free auditing to enable upstream detection (e.g. auditing before diffusion denoising completion \cite{yuan2025lurks, li2025detect}.) 
    \item Harden models to white-box attacks using natural-language safety specifications \cite{guan2024deliberative}.
    \item Develop training data extraction techniques that do not rely on direct prompting or model training
    \item Explore concept fusion via MI to identify and downweight neurons enabling adult-child combinations. 
\end{itemize}}
\end{mybox}

\vspace{.05in}
AI Providers have unique visibility into how offenders are misusing their platforms. They should prioritize efforts to use and share this data and other related metadata available to strengthen safeguards, including efforts such as: 

\begin{mybox}
{\textbf{AI Providers}}{
\begin{itemize}[itemsep=0ex,leftmargin=*]
\item Join data sharing programs (e.g. \cite{TechCoalitionLantern}) with trusted organizations to broaden access to adversarial prompts across platforms.
\item Deploy metadata and other signals to detect policy-violating models. Resource teams to meet the volume of models uploaded/downloaded.
\item Partner with external teams pre-deployment for model evaluation, sharing platform-specific offender behavior data for effective assessment
\end{itemize}}
\end{mybox}

\vspace{.05in}
Policymakers are key to establishing broader systems and structures for building public trust in safety technologies used to safeguard generative models. They also play a critical role in drawing clear lines in the sand that allow for scalable, preventative efforts.

\begin{mybox}[after skip=0pt]
{\textbf{Policymakers}}{
\begin{itemize}[itemsep=0ex,leftmargin=*] 
\item Resource institutions like NIST to establish pathways for public benchmarking of safety tech.
\item Pass legislation creating liability for intentional development or distribution of models built to produce AIG-CSAM or “nudify” pictures of children.
\item Require that developers disclose whether they conducted CSAM filtering on their training data (e.g. \cite{EU_GPAI_Code_2025}, \cite{eSafetytransparencyreport0825})
\end{itemize}
}
\end{mybox}

\vspace{.2in}
\section{Safe Model Maintenance}
\label{sec:maintain}
Finally, model monitoring and maintenance is necessary to address emerging and evolving threats to children, maintain efficacy against an evolving technology stack, and reflect the broad nature of the ecosystem (where interweaving technology providers and systems may inadvertently reinforce harms, or create new harms \cite{thorn2024evolving}). In particular, open source models are highly vulnerable to abliteration: fine-tuning to remove safety guardrails \cite{henderson2024safety}, allowing offenders to enhance CSAM generation or ``nudify'' minors. Technology aggregators (model hosting platforms, app stores, search engines, AI character platforms, etc.) enable easy discovery of these harmful models \cite{stokelwalker2025chatbots,Thorn_DeepfakeNudes_2024}. 



\subsection {Current Approaches}
Current industry safety solutions for detecting harmful models are primarily reactive---removing flagged models rather than evaluating 
before providing access. While ``nudifying'' applications can be easily discovered via simple text matching searches \cite{gibson2025analyzing}, interventions to delist or suppress such sites and resources are largely lacking.
Those that do incorporate detection strategies report using model hashlists (blocking uploads of files that are on hashlists of known CSAM optimized models~\cite{hawkins2025deepfakes,thorn2025safetybydesignupdate}) or similar efforts such as removing advertisements of ``nudifying'' applications \cite{meta2025nudify}. 
Industry further reports efforts to maintain the quality of their own detection technologies, include child safety policies for each of their services, 
and collaborate with child safety organizations \cite{thorn2025safetybydesignupdate}. 

\subsection{Open Problems}
\begin{pboxchipped}
{\textbf{Open Problem C1: Identifying abliterated models}}
{\chipEval\ \chipAdv}
{How can we identify models and services that have been optimized for CSAM and ``nudification''?}
\end{pboxchipped}

With thousands of models, apps, and services created and uploaded daily, rapid identification of those optimized for harmful purposes remains a critical gap. 

\emph{Existing work.} Model fingerprinting uses adversarial attacks to compare outputs between original and suspected stolen models for IP protection \cite{guan2022you}. Model diffing exploits mechanistic differences between base and fine-tuned models to identify model-specific concepts \cite{anthropic2024modeldiff, minder2026overcoming}. AIG-CSAM model hashlists can be built using cryptographic hashing.

\emph{Limitations.} Model fingerprinting requires insight into the ``original'' model, which is challenging to determine for models optimized by malicious actors. Model diffing is not well explored for text-to-image models, particularly for LoRA fine-tuning (often used by CSAM perpetrators). Cryptographic hashing detects exact model replicas but not minor modifications. Further, sourcing such hashlists can require accessing Tor onion sites dedicated to child abuse, which comes with significant legal and wellness risks. 

\begin{pboxchipped}
{\textbf{Open Problem C2: Robust unlearning}}
{\chipData\ \chipAdv}
{How can we reliably erase the concept of CSAM from generative models?}
\end{pboxchipped}

CSAM might be found in training data after a model is deployed; concept fusion opens up other pathways for CSAM generation. 
The gold standard is to retrain the model from scratch without harmful data, but retraining can be prohibitively expensive and time-consuming \cite{cooper2024machine}. If the model was built by a different developer than the person who discovered the issue, engaging the original developer to conduct a full re-train may be challenging. 


\emph{Limitations.} Existing work has extensively explored \textit{approximate machine unlearning} or \textit{concept erasure} for text-to-image models, to remove knowledge with only a textual description of the harmful concept \cite{gandikota2023erasing, lu2024mace}. However, these methods are vulnerable to adversarial prompts \cite{zhang2024defensive} and only provide probabilistic guarantees that a model no longer retains certain knowledge. CSAM requires \textit{exact unlearning}, providing strict guarantees that CSAM images have no effect on model output \cite{cooper2024machine, bourtoule2021sisa}. 

\begin{pboxchipped}
{\textbf{Open Problem C3: Protecting user imagery}}
{\chipEval\ \chipAdv}
{How do we proactively protect users’ imagery from unwanted AI-generated manipulation?}
\end{pboxchipped}

Solutions at the model level to prevent adversarial optimization are necessary, but do not afford end users (or platforms hosting user generated content) \textit{agency} to proactively protect their own content from unwanted AI-manipulation.

\emph{Existing work.} Research exploring image immunization \cite{salman2023raising} involves injecting imperceptible perturbations into the image, such that image editing software fails to successfully edit the image. Some recent efforts specifically focus on protecting children's imagery \cite{afp_monash_silverer_2025}. In the IP protection space, similar solutions have been proposed to disrput a model's ability to mimic the style of particular artists \cite{shan2023glaze}. 

\emph{Limitations.} Some research indicates that these image perturbation strategies are not robust to simple attacks such as image upscaling \cite{honig2024adversarial}. Further, solutions to protect video imagery are lacking. Effectively evaluating these techniques ability to protect children's imagery from ``nudification'' requires attempting to generate such imagery, which has ethical and legal implications.

\begin{pboxchipped}
{\textbf{Open Problem C4: Securing AI agents}}
{\chipData\ \chipAdv\ \chipWell}
{How can we prevent the misuse of AI agents and code generation to facilitate child sexual abuse?}
\end{pboxchipped}

Although criminal actors do not yet appear to be adopting automated code generation and AI agents for child sexual exploitation, the tools are already being used in other criminal enterprises, such as malware creation \cite{Anthropic_ThreatIntel_2025} and building hidden webcam recording software \cite{gtig_adversarial_misuse_2025}. AI agents capable of relationship building could enable sexual extortion schemes; code generation could automate the creation of ``nudifying'' software, even if prebuilt nudifying apps are banned.

\emph{Existing work.} Safety for agentic systems is an emerging field \cite{Song_2024_AgentSafety}. Current work focuses on early detection and prevention of misuse. As with traditional red teaming and filtering, most solutions rely on training models using examples of prior misuse, such as AI-generated malware \cite{deepmind_codemender_2025, OpenAI_IntroducingAardvark_2025, anthropic_claude_code_security_2025}. Such examples are legal to develop and train on, and leading labs actively build this data. 

\emph{Limitations.} Developing ``nudification'' code or sexual extortion prompts is more ethically ambiguous than malware, raising data legality and researcher wellness concerns similar to directly red teaming for CSAM. Robust evidence that these emerging technologies are being misused in child safety contexts may be a prerequisite for industry prioritization, but obtaining the visibility needed to build such evidence remains challenging.

\textit{We discuss additional open problems in model maintenance, such as safeguard assessments, in App.~\ref{app:maintenance}.}



\subsection{Call to Action}

Interventions for safe model maintenance are unique in that they can be both cross-platform and model specific, requiring research that covers broad surface areas for harm.

\begin{mybox}
{\textbf{Research Directions}}{
\begin{itemize}[wide, labelwidth=0pt, labelindent=0pt, topsep=0pt, partopsep=0pt, parsep=0pt, itemsep=0ex]
    \item Build solutions to detect ``nudifying'' applications.
    \item Develop model hashing techniques that are robust to minor modifications of the model.
    \item Explore strategies that provide strong guarantees when remediating/unlearning harmful models. 
    \item Establish image/video protection techniques for chidren's imagery, using proxy data and concepts.
\end{itemize}}
\end{mybox}

\vspace{-.05in}
AI Providers are already positioned to engage in both model specific and cross-platform safeguarding efforts:
\vspace{-.05in}

\begin{mybox}
{\textbf{AI Providers}}{
\begin{itemize}[wide, labelwidth=0pt, labelindent=0pt, topsep=0pt, partopsep=0pt, parsep=0pt, itemsep=0ex]
\item Partner with cross-industry child safety organizations (e.g. the Tech Coalition) to enable cross-platform monitoring and measurement.
\item Partner with user-generated content platforms to evaluate strategies for protecting user imagery from unwanted AI-generated manipulation originating from your foundation models
\item Model hosting platforms jointly establish cross-industry consistency in policies and enforcement for identifying \& removing harmful third-party models 
\end{itemize}}
\end{mybox}

\vspace{-.05in}
Policymakers can create incentives and pathways such that stakeholders invest in cross-platform safety:
\vspace{-.05in}

\begin{mybox}
{\textbf{Policymakers}}{

\begin{itemize}[wide, labelwidth=0pt, labelindent=0pt, topsep=0pt, partopsep=0pt, parsep=0pt, itemsep=0ex] 
\item Task regulators to review “nudifying” services (websites, apps, etc.) for unfair, deceptive and fraudulent business practices that promote AIG-CSAM creation and distribution.
\item Mandate child safety specific impact and risk assessments for AI systems before deployment. 
\item Direct regulators to assess platform (e.g. model hosting companies, search engines and app stores) strategies for preventing the distribution of models, apps and services that enable CSEA.
\end{itemize}
}
\end{mybox}

\section{Alternative Views}
\label{sec:alternate}

The most immediate alternative perspective is  that child safety does not require fundamentally new AI safety approaches, but rather stronger or more consistently applied variants of existing approaches. Indeed, the core technical challenges (e.g., dataset governance, robust content filtering, adversarial testing, and post-deployment monitoring) are not unique to child safety and are broadly relevant to problems like disinformation, non-consensual imagery, and extremist content. Our work argues that despite the seeming similarities to these problems, the legal and ethical restrictions surrounding AIG-CSAM present non-trivial additional hurdles to existing approaches, necessitating in some cases entirely new techniques. While the problems identified are necessary to reduce the harms of AIG-CSAM, we believe many of them could prove beneficial for these other, less regulated domains.

Another view is that the problem of AIG-CSAM simply cannot be solved, or that we should instead focus AI safety efforts on other, existential risks~\cite{hendrycks2023overview}. Others may go further to suggest that AIG-CSAM is not harmful or is at least preferable to non-AI generated CSAM. Here, we note the significant evidence to the contrary~\cite{ciardha2025ai}, and advocate for a balanced position to invest efforts in multiple safety areas, especially as we could potentially make immediate headway on reducing the harms of AIG-CSAM during a time when reports are increasing sharply~\cite{2025report,iwfblog,iwf2024}.

Finally, one may question whether technical AI safety interventions should be the primary locus of response at all. We could instead emphasize social, legal, and preventive measures such as digital literacy education for children and parents, victim support services, updated criminal law and international cooperation, and platform liability regimes that deter harmful deployment. From this standpoint, overemphasis on building new AI-specific safeguards may create a false sense of technological solvability while underinvesting in root causes. Others may go further to argue that policy solutions would eliminate the need for developing new technical approaches altogether, reducing the legal barriers to applying existing safety approaches. 
Our position reflects that policy solutions must come paired with, and be informed by, technical advancements.


\section{Conclusion, Limitations \& Future Work}
\label{sec:conclusion}
In this work we argue that current safety techniques assume data accessibility, transparency, and evaluation practices that are in many cases incompatible with ethical and legal constraints surrounding CSAM. We then propose targeted recommendations and a set of \textit{15 open technical problems} that can improve child safety in the AI ecosystem.

We note limitations and areas for future work in our study. 
We focus on \textit{AI-generated imagery}. However, exploitation spans multiple modalities, 
including text sexualizing minors \cite{graphika25}, 
generative chatbots using celebrity voices for sexual conversations with children's accounts \cite{horwitz2025}, and video deepfakes of child sexual abuse \cite{kamachee2025video}. 
While some of the issues identified in this work may also translate to other modalities, 
we leave the identification of technical open problems in non-image modalities as future work. 

We also note that the risks and open problems identified throughout are primarily from a \textit{U.S.-centric perspective}. The policy institutions, regulatory bodies, and frameworks referenced are also primarily U.S.-centric. Global regulation in this space is actively evolving \cite{politico_uk_nudification_ban_2025, bbc_nudify_fine_2025}, and those evolutions may add additional nuance and context to the content in this work.  
A short discussion of policy tradeoffs is provided in Appendix \ref{app:policytradeoffs}, but exact policy implementations are left for future work and discussion with policymakers. 

Moving forward, measuring and reducing the utility cost of existing safety methods is critical for adoption. At present, most AI-generated abusive material and models optimized for AIG-CSAM derive from open-source models. As such, the most pressing open problems are those that address the adversarial misuse of open-source models.  Resilience to harmful fine-tuning (A3) is imperative, as perpetrators actively use GUI-based LoRA tools to abliterate models for AIG-CSAM generation. Other urgent problems include watermark robustness (A5), moderating the hobbyist ecosystem (B5), detecting abliterated models (C1), and robust unlearning (C2); we direct readers to \cite{casper2025open} on broader open-source safety risks. Establishing collaborative venues (e.g., conference workshops) that foster communication between stakeholders could accelerate progress on these open problems.

\clearpage

\section*{Acknowledgments}

We thank Steven Wu for his contributions in early planning for this work, and Stephen Casper for his feedback on later drafts. We also thank Tim O'Gorman, James Williams, Jim Pitkow, and Melissa Stroebel from Thorn for their reviews and comments on the paper.


\bibliographystyle{icml2026}
\bibliography{refs}

@article{georgiev2023journey,
  title={The journey, not the destination: How data guides diffusion models},
  author={Georgiev, Kristian and Vendrow, Joshua and Salman, Hadi and Park, Sung Min and Madry, Aleksander},
  journal={arXiv preprint arXiv:2312.06205},
  year={2023}
}

@inproceedings{lin2025diffusion,
  title={Diffusion attribution score: Evaluating training data influence in diffusion models},
  author={Lin, Jinxu and Tao, Linwei and Dong, Minjing and Xu, Chang},
  booktitle={International Conference on Learning Representations},
  volume={2025},
  pages={22962--22989},
  year={2025}
}

@inproceedings{zheng2024intriguing,
  title={Intriguing properties of data attribution on diffusion models},
  author={Zheng, Xiaosen and Pang, Tianyu and Du, Chao and Jiang, Jing and Lin, Min},
  booktitle={International Conference on Learning Representations},
  volume={2024},
  pages={18417--18452},
  year={2024}
}

@article{tinaz2026emergence,
  title={Emergence and evolution of interpretable concepts in diffusion models},
  author={Tinaz, Berk and Fabian, Zalan and Soltanolkotabi, Mahdi},
  journal={Advances in Neural Information Processing Systems},
  year={2026}
}

@article{surkov2024one,
  title={One-Step is Enough: Sparse Autoencoders for Text-to-Image Diffusion Models},
  author={Surkov, Viacheslav and Wendler, Chris and Mari, Antonio and Terekhov, Mikhail and Deschenaux, Justin and West, Robert and Gulcehre, Caglar and Bau, David},
  journal={Advances in Neural Information Processing Systems},
  year={2025}
}

@misc{hipaa1996congress,
  author       = {{U.S. Congress}},
  title        = {Health Insurance Portability and Accountability Act of 1996},
  howpublished = {Public Law 104--191, 110 Stat. 1936},
  year         = {1996},
  month        = aug,
  note         = {Accessed: 2026-05-16},
  url          = {https://www.govinfo.gov/app/details/PLAW-104publ191}
}

@misc{natsec1995potus,
  author       = {Clinton, William J.},
  title        = {Executive Order 12968: Access to Classified Information},
  howpublished = {60 Fed. Reg. 40245},
  year         = {1995},
  month        = aug,
  day          = {2},
  note         = {Accessed: 2026-05-16},
  url          = {https://www.archives.gov/isoo/policy-documents/eo-12968.html}
}

@techreport{csam2023icmec,
  author      = {{International Centre for Missing \& Exploited Children}},
  title       = {Child Sexual Abuse Material: Model Legislation \& Global Review},
  edition     = {10},
  institution = {International Centre for Missing \& Exploited Children},
  year        = {2023},
  month       = oct,
  address     = {Alexandria, VA},
  url         = {https://cdn.icmec.org/wp-content/uploads/2023/10/CSAM-Model-Legislation_10th-Ed-Oct-2023.pdf},
  note        = {Accessed: 2026-05-16}
}

@misc{citizens2026doj,
  author       = {{U.S. Department of Justice}},
  title        = {Citizen's Guide to {U.S.} Federal Law on Child Pornography},
  howpublished = {\url{https://www.justice.gov/criminal/criminal-ceos/citizens-guide-us-federal-law-child-pornography}},
  year = {2023},
  note         = {Accessed: 2026-05-16}
}

@inproceedings{das2020fast,
  title={Fast, accurate, and healthier: Interactive blurring helps moderators reduce exposure to harmful content},
  author={Das, Anubrata and Dang, Brandon and Lease, Matthew},
  booktitle={AAAI Conference on Human Computation and Crowdsourcing},
  year={2020}
}

@article{casper2025open,
  title={Open technical problems in open-weight ai model risk management},
  author={Casper, Stephen and O’Brien, Kyle and Longpre, Shayne and Seger, Elizabeth and Klyman, Kevin and Bommasani, Rishi and Nrusimha, Aniruddha and Shumailov, Ilia and Mindermann, S{\"o}ren and Basart, Steven and others},
  journal={Social Science Research Network},
  year={2025}
}

@article{kamachee2025video,
  title={Video Deepfake Abuse: How Company Choices Predictably Shape Misuse Patterns},
  author={Kamachee, Max and Casper, Stephen and Ding, Michelle L and Yew, Rui-Jie and Reuel, Anka and Biderman, Stella and Hadfield-Menell, Dylan},
  journal={arXiv preprint arXiv:2512.11815},
  year={2025}
}

@inproceedings{nichol2021glide,
  title={Glide: Towards photorealistic image generation and editing with text-guided diffusion models},
  author={Nichol, Alex and Dhariwal, Prafulla and Ramesh, Aditya and Shyam, Pranav and Mishkin, Pamela and McGrew, Bob and Sutskever, Ilya and Chen, Mark},
  booktitle={International Conference on Machine Learning},
  year={2022}
}

@article{ding2025malicious,
  title={The Malicious Technical Ecosystem: Exposing Limitations in Technical Governance of AI-Generated Non-Consensual Intimate Images of Adults},
  author={Ding, Michelle L and Suresh, Harini},
  journal={arXiv preprint arXiv:2504.17663},
  year={2025}
}

@article{bbc_nudify_fine_2025,
  author       = {McMahon, Liv},
  title        = {Deepfake 'Nudify' Site Fined £55,000 Over Lack of Age Checks},
  journal      = {BBC News},
  year         = {2025},
  month        = nov,
  url          = {https://www.bbc.com/news/articles/cn8xq677l9xo},
}

@article{politico_uk_nudification_ban_2025,
  author       = {Clifton, Mizy and Bristow, Tom},
  title        = {{UK} Set to Ban Deepfake Nudification Apps},
  journal      = {Politico Europe},
  year         = {2025},
  url          = {https://www.politico.eu/article/uk-set-to-ban-deepfake-nudification-apps-vawg/},
  note         = {Accessed November 2025}
}

@misc{anthropic_claude_code_security_2025,
  author       = {{Anthropic}},
  title        = {Automate Security Reviews with {Claude Code}},
  year         = {2025},
  month        = aug,
  day          = {6},
  howpublished = {Claude Blog},
  url          = {https://www.claude.com/blog/automate-security-reviews-with-claude-code},
  note         = {Accessed November 2025}
}

@article{OpenAI_IntroducingAardvark_2025,
  author  = {{OpenAI}},
  title   = {Introducing Aardvark: OpenAI’s agentic security researcher},
  journal = {OpenAI},
  year    = {2025},
  url     = {https://openai.com/index/introducing-aardvark/},
  urldate = {2025-11-26}
}

@misc{deepmind_codemender_2025,
  author       = {Popa, Raluca Ada and Flynn, Four},
  title        = {Introducing {CodeMender}: An {AI} Agent for Code Security},
  year         = {2025},
  howpublished = {Google DeepMind Blog},
  url          = {https://deepmind.google/discover/blog/introducing-codemender-an-ai-agent-for-code-security/},
}

@misc{Song_2024_AgentSafety,
  author       = {Song, Dawn},
  title        = {Towards Building Safe \& Trustworthy AI Agents and A Path for Science- and Evidence-based AI Policy},
  year         = {2024},
  howpublished = {Lecture slides for CS294/CS194-196: Large Language Model Agents, UC Berkeley},
  url          = {https://rdi.berkeley.edu/llm-agents/assets/dawn-agent-safety.pdf},
}

@misc{gtig_adversarial_misuse_2025,
  author       = {{Google Threat Intelligence Group}},
  title        = {Adversarial Misuse of Generative {AI}},
  year         = {2025},
  howpublished = {Google Cloud Blog},
  url          = {https://cloud.google.com/blog/topics/threat-intelligence/adversarial-misuse-generative-ai},
}

@techreport{Anthropic_ThreatIntel_2025,
  author      = {{Anthropic Threat Intelligence Team}},
  title       = {Threat Intelligence Report: August 2025},
  institution = {Anthropic},
  year        = {2025},
  month       = aug,
  url         = {https://www-cdn.anthropic.com/b2a76c6f6992465c09a6f2fce282f6c0cea8c200.pdf},
  urldate     = {2025-11-26}
}

@misc{afp_monash_silverer_2025,
  author       = {{Australian Federal Police} and {Monash University}},
  title        = {{AFP} and {Monash University} Poison Data to Combat {AI}-Generative Crime},
  year         = {2025},
  howpublished = {Joint media release},
  url          = {https://www.afp.gov.au/news-centre/media-release/afp-and-monash-university-poison-data-combat-ai-generative-crime},
  note         = {Accessed November 2025}
}

@article{guan2022you,
  title={Are you stealing my model? sample correlation for fingerprinting deep neural networks},
  author={Guan, Jiyang and Liang, Jian and He, Ran},
  journal={Advances in Neural Information Processing Systems},
  year={2022}
}

@article{doerfler2021m,
  title={``{I'm} a Professor, which isn't usually a dangerous job''': Internet-facilitated Harassment and Its Impact on Researchers},
  author={Doerfler, Periwinkle and Forte, Andrea and De Cristofaro, Emiliano and Stringhini, Gianluca and Blackburn, Jeremy and McCoy, Damon},
  journal={Proceedings of the ACM on Human-Computer Interaction},
  year={2021},
}

@article{koch2014joining,
  title={Joining a smartphone ecosystem: Application developers’ motivations and decision criteria},
  author={Koch, Stefan and Kerschbaum, Markus},
  journal={Information and Software Technology},
  volume={56},
  number={11},
  pages={1423--1435},
  year={2014},
  publisher={Elsevier}
}

@article{patterson2022systematic,
  title={A systematic review of the education and awareness interventions to prevent online child sexual abuse},
  author={Patterson, Anastasia and Ryckman, Leah and Guerra, Crist{\'o}bal},
  journal={Journal of Child \& Adolescent Trauma},
  year={2022},
  publisher={Springer}
}

@article{wurtele2010partnering,
  title={Partnering with parents to prevent childhood sexual abuse},
  author={Wurtele, Sandy K and Kenny, Maureen C},
  journal={Child Abuse Review: Journal of the British Association for the Study and Prevention of Child Abuse and Neglect},
  year={2010},
  publisher={Wiley Online Library}
}

@inproceedings{nunes2022using,
  title={Using model cards for ethical reflection: a qualitative exploration},
  author={Nunes, Jos{\'e} Luiz and Barbosa, Gabriel DJ and De Souza, Clarisse Sieckenius and Lopes, Helio and Barbosa, Simone DJ},
  booktitle={Proceedings of the 21st Brazilian Symposium on Human Factors in Computing Systems},
  year={2022}
}

@inproceedings{crisan2022interactive,
  title={Interactive model cards: A human-centered approach to model documentation},
  author={Crisan, Anamaria and Drouhard, Margaret and Vig, Jesse and Rajani, Nazneen},
  booktitle={ACM Conference on Fairness, Accountability, and Transparency},
  year={2022}
}

@inproceedings{mitchell2019model,
  title={Model cards for model reporting},
  author={Mitchell, Margaret and Wu, Simone and Zaldivar, Andrew and Barnes, Parker and Vasserman, Lucy and Hutchinson, Ben and Spitzer, Elena and Raji, Inioluwa Deborah and Gebru, Timnit},
  booktitle={Proceedings of the conference on fairness, accountability, and transparency},
  year={2019}
}

@article{rauh2024gaps, 
    title={Gaps in the Safety Evaluation of Generative AI}, 
    journal={Proceedings of the AAAI/ACM Conference on AI, Ethics, and Society}, 
    author={Rauh, Maribeth and Marchal, Nahema and Manzini, Arianna and Hendricks, Lisa Anne and Comanescu, Ramona and Akbulut, Canfer and Stepleton, Tom and Mateos-Garcia, Juan and Bergman, Stevie and Kay, Jackie and Griffin, Conor and Bariach, Ben and Gabriel, Iason and Rieser, Verena and Isaac, William and Weidinger, Laura}, 
    year={2024}, 
}

@article{HarrisWillner_StableDiffusionCSAM_2024,
  author  = {Harris, David Evan and Willner, Dave},
  title   = {Was an AI Image Generator Taken Down for Making Child Porn?},
  journal = {IEEE Spectrum},
  year    = {2024},
  url     = {https://spectrum.ieee.org/stable-diffusion},
  urldate = {2025-11-26}
}

@article{griffin2025ethical,
  title={The ethical wisdom of AI developers},
  author={Griffin, Tricia A and Green, Brian P and Welie, Jos VM},
  journal={AI and Ethics},
  volume={5},
  number={2},
  pages={1087--1097},
  year={2025},
  publisher={Springer}
}

@article{guardian2023aitoll,
  title        = {Kenyan Moderators Decry Toll of Training of AI Models},
  author       = {Rowe, Niamh},
  journal      = {The Guardian},
  year         = {2023},
  url          = {https://www.theguardian.com/technology/2023/aug/02/ai-chatbot-training-human-toll-content-moderator-meta-openai},
}

@article{rimer2025once,
  title={“Once you see it you can't unsee it”: Law enforcement trauma and immersion in child sexual abuse material},
  author={Rimer, Jonah R and Brown, Shannon and Martin, Jennifer and Slane, Andrea},
  journal={Child Protection and Practice},
  year={2025},
  publisher={Elsevier}
}

@techreport{Thorn_PAI_CaseStudy_2024,
  author      = {{Thorn}},
  title       = {Mitigating the Risk of Generative AI Models Creating Child Sexual Abuse Materials},
  institution = {Partnership on AI},
  year        = {2024},
  month       = nov,
  type        = {Case Study},
  url         = {https://partnershiponai.org/wp-content/uploads/2024/11/case-study-thorn.pdf},
}

@article{spence2023psychological,
  title={The psychological impacts of content moderation on content moderators: A qualitative study},
  author={Spence, Ruth and Bifulco, Antonia and Bradbury, Paula and Martellozzo, Elena and DeMarco, Jeffrey},
  journal={Cyberpsychology: Journal of Psychosocial Research on Cyberspace},
  year={2023}
}

@misc{ostris2024aitoolkit,
  author       = {Ostris},
  title        = {{AI Toolkit}: The Ultimate Training Toolkit for Finetuning Diffusion Models},
  year         = {2024},
  url          = {https://github.com/ostris/ai-toolkit},
  version      = {1.0},
}

@misc{gdpreu2018whatgdpr,
  title        = {What is {GDPR}, the {EU}'s New Data Protection Law?},
  author       = {{GDPR.eu}},
  year         = {2018},
  url          = {https://gdpr.eu/what-is-gdpr/},
  urldate      = {2025-11-26},
  organization = {Proton AG},
}

@inproceedings{huunlearning,
  title={Unlearning or Obfuscating? Jogging the Memory of Unlearned LLMs via Benign Relearning},
  author={Hu, Shengyuan and Fu, Yiwei and Wu, Steven and Smith, Virginia},
  booktitle={International Conference on Learning Representations},
  year={2025}
}

@article{cretu2025evaluating,
  title={Evaluating Concept Filtering Defenses against Child Sexual Abuse Material Generation by Text-to-Image Models},
  author={Cretu, Ana-Maria and Kireev, Klim and Abdalla, Amro and Obinna, Wisdom and Meier, Raphael and Bargal, Sarah Adel and Redmiles, Elissa M and Troncoso, Carmela},
  journal={arXiv preprint arXiv:2512.05707},
  year={2025}
}

@misc{C2PA_Org,
  author       = {{Coalition for Content Provenance and Authenticity}},
  title        = {C2PA: Advancing Digital Content Transparency and Authenticity},
  howpublished = {\url{https://c2pa.org/}},
  year         = {2024},
  note         = {Coalition for Content Provenance and Authenticity},
  urldate      = {2025-11-26}
}

@misc{FTC_COPPA,
  author       = {{Federal Trade Commission}},
  title        = {Children's Online Privacy Protection Rule ("COPPA")},
  url          = {https://www.ftc.gov/legal-library/browse/rules/childrens-online-privacy-protection-rule-coppa},
  note         = {Accessed: 2025-11-26},
  organization = {Federal Trade Commission},
  year         = {2025}
}

@article{stokelwalker2025chatbots,
  title        = {Thousands of Pedophiles Are Using Jail-Broken {AI} Character Chatbots to Roleplay Sexually Assaulting Minors},
  author       = {Stokel-Walker, Chris},
  journal      = {Fast Company},
  year         = {2025},
  url          = {https://www.fastcompany.com/91290478/graphika-report-ai-chatbots-role-playing-sex-with-minors},
}

@misc{meta2025nudify,
  title        = {Taking Action Against Nudify Apps},
  author       = {{Meta}},
  year         = {2025},
  month        = jun,
  organization = {Meta Platforms, Inc.},
  url          = {https://about.fb.com/news/2025/06/taking-action-against-nudify-apps/},
  urldate      = {2025-06-10},
}

@misc{Thorn_DeepfakeNudes_2024,
  author       = {{Thorn}},
  title        = {Deepfake Nudes \& Young People: Navigating a New Frontier in Technology-facilitated Nonconsensual Sexual Abuse and Exploitation},
  year         = {2024},
  institution  = {Thorn},
  howpublished = {\url{https://www.thorn.org/research/library/deepfake-nudes-and-young-people/}},
  urldate      = {2025-11-26}
}

@techreport{thorn2024evolving,
  title        = {Evolving Technologies Horizon Scan: A Review of Technologies Carrying Notable Risk and Opportunity in the Fight Against Technology-Facilitated Child Sexual Exploitation},
  author       = {{Thorn} and {WeProtect Global Alliance}},
  year         = {2024},
  month        = dec,
  institution  = {Thorn and WeProtect Global Alliance},
  type         = {Technical Report},
  url          = {https://info.thorn.org/hubfs/Research/Thorn_x_WPGA_EvolvingTechnologies_Dec2024.pdf},
}

@misc{EU_GPAI_Code_2025,
  title        = {The General-Purpose AI Code of Practice},
  author       = {{European Commission}},
  howpublished = {\url{https://digital-strategy.ec.europa.eu/en/policies/contents-code-gpai}},
  note         = {Shaping Europe’s digital future},
  year         = {2025},
  month        = {July},
  urldate      = {2025-11-26}
}

@misc{TechCoalitionLantern,
  author       = {{Tech Coalition}},
  title        = {Lantern: Advancing Child Safety Through Signal Sharing},
  year         = {2025},
  url          = {https://technologycoalition.org/programs/lantern/},
  note         = {Accessed: 2025-11-26},
  organization = {Technology Coalition}
}

@inproceedings{liu2015deep,
  title={Deep learning face attributes in the wild},
  author={Liu, Ziwei and Luo, Ping and Wang, Xiaogang and Tang, Xiaoou},
  booktitle={IEEE/CVF International Conference on Computer Vision},
  year={2015}
}

@article{ren2024safetywashing,
  title={Safetywashing: Do AI Safety Benchmarks Actually Measure Safety Progress?},
  author={Richard Ren and Steven Basart and Adam Khoja and Alice Gatti and Long Phan and Xuwang Yin and Mantas Mazeika and Alexander Pan and Gabriel Mukobi and Ryan H. Kim and Stephen Fitz and Dan Hendrycks},
  journal={Advances in Neural Information Processing Systems},
  year={2024}
}

@inproceedings{shan2023glaze,
  title={Glaze: Protecting artists from style mimicry by Text-to-Image models},
  author={Shan, Shawn and Cryan, Jenna and Wenger, Emily and Zheng, Haitao and Hanocka, Rana and Zhao, Ben Y},
  booktitle={USENIX Security Symposium},
  year={2023}
}

@article{wang2024disentangled,
  title={Disentangled representation learning},
  author={Wang, Xin and Chen, Hong and Tang, Si'ao and Wu, Zihao and Zhu, Wenwu},
  journal={IEEE Transactions on Pattern Analysis and Machine Intelligence},
  year={2024},
  publisher={IEEE}
}

@article{espinosa2022concept,
  title={Concept embedding models: Beyond the accuracy-explainability trade-off},
  author={Zarlenga, Mateo Espinosa and Barbiero, Pietro and Ciravegna, Gabriele and Marra, Giuseppe and Giannini, Francesco and Diligenti, Michelangelo and Shams, Zohreh and Precioso, Frederic and Melacci, Stefano and Weller, Adrian and Lió, Pietro and Jamnik, Mateja},
  journal={Advances in Neural Information Processing Systems},
  year={2022}
}

@article{vidgen2024introducing,
  title={Introducing v0. 5 of the ai safety benchmark from mlcommons},
  author={Vidgen, Bertie and Agrawal, Adarsh and Ahmed, Ahmed M and Akinwande, Victor and Al-Nuaimi, Namir and Alfaraj, Najla and Alhajjar, Elie and Aroyo, Lora and Bavalatti, Trupti and Bartolo, Max and others},
  journal={arXiv preprint arXiv:2404.12241},
  year={2024}
}

@article{ahmad2025openai,
  title={OpenAI's Approach to External Red Teaming for AI Models and Systems},
  author={Ahmad, Lama and Agarwal, Sandhini and Lampe, Michael and Mishkin, Pamela},
  journal={arXiv preprint arXiv:2503.16431},
  year={2025}
}

@inproceedings{zhang2024human,
  title={The human factor in ai red teaming: Perspectives from social and collaborative computing},
  author={Zhang, Alice Qian and Shaw, Ryland and Anthis, Jacy Reese and Milton, Ashlee and Tseng, Emily and Suh, Jina and Ahmad, Lama and Kumar, Ram Shankar Siva and Posada, Julian and Shestakofsky, Benjamin and others},
  booktitle={Conference on Computer-Supported Cooperative Work and Social Computing},
  year={2024}
}

@article{ciardha2025ai,
  title={AI Generated Child Sexual Abuse Material--What's the Harm?},
  author={Ciardha, Caoilte {\'O} and Buckley, John and Portnoff, Rebecca S},
  journal={arXiv preprint arXiv:2510.02978},
  year={2025}
}

@inproceedings{honig2024adversarial,
  title={Adversarial perturbations cannot reliably protect artists from generative ai},
  author={H{\"o}nig, Robert and Rando, Javier and Carlini, Nicholas and Tram{\`e}r, Florian},
  booktitle={International Conference on Learning Representations},
  year={2025}
}

@article{salman2023raising,
  title={Raising the cost of malicious ai-powered image editing},
  author={Salman, Hadi and Khaddaj, Alaa and Leclerc, Guillaume and Ilyas, Andrew and Madry, Aleksander},
  journal={International Conference on Machine Learning},
  year={2023}
}

@techreport{thorn2025safetybydesignupdate,
  author      = {Portnoff, Rebecca and Simpson, Michael},
  title       = {Safety by Design: Annual Progress Report (Report \#4: April 2024 to April 2025)},
  institution = {Thorn},
  year        = {2025},
  url         = {https://info.thorn.org/hubfs/Thorn_SafetyByDesign_AnnualProgressReport_April2024-April2025.pdf},
  note        = {Design \& Publication by Yena Lee, Cassie Coccaro, and Justus Hyatt}
}

@inproceedings{hawkins2025deepfakes,
author = {Hawkins, Will and Mittelstadt, Brent and Russell, Chris},
title = {Deepfakes on Demand: The rise of accessible non-consensual deepfake image generators},
year = {2025},
booktitle = {ACM Conference on Fairness, Accountability, and Transparency}
}

@inproceedings{gibson2025analyzing,
  title={Analyzing the {AI} Nudification Application Ecosystem},
  author={Gibson, Cassidy and Olszewski, Daniel and Brigham, Natalie Grace and Crowder, Anna and Butler, Kevin RB and Traynor, Patrick and Redmiles, Elissa M and Kohno, Tadayoshi},
  booktitle={USENIX Security Symposium},
  year={2025}
}

@inproceedings{carlini2023extracting,
  title={Extracting training data from diffusion models},
  author={Carlini, Nicolas and Hayes, Jamie and Nasr, Milad and Jagielski, Matthew and Sehwag, Vikash and Tramer, Florian and Balle, Borja and Ippolito, Daphne and Wallace, Eric},
  booktitle={USENIX Security Symposium},
  year={2023}
}

@inproceedings{bourtoule2021sisa,
  title={Machine unlearning},
  author={Bourtoule, Lucas and Chandrasekaran, Varun and Choquette-Choo, Christopher A and Jia, Hengrui and Travers, Adelin and Zhang, Baiwu and Lie, David and Papernot, Nicolas},
  booktitle={IEEE Symposium on Security and Privacy},
  year={2021},
}

@inproceedings{cooper2024machine,
  title={Machine Unlearning Doesn't Do What You Think: Lessons for Generative AI Policy, Research, and Practice},
  author={Cooper, A Feder and Choquette-Choo, Christopher A and Bogen, Miranda and Jagielski, Matthew and Filippova, Katja and Liu, Ken Ziyu and Chouldechova, Alexandra and Hayes, Jamie and Huang, Yangsibo and Mireshghallah, Niloofar and others},
  booktitle={Advances in Neural Information Processing Systems, Position Paper Track},
  year={2025}
}

@article{gowal2025synthid,
  title={SynthID-Image: Image watermarking at internet scale},
  author={Gowal, Sven and Bunel, Rudy and Stimberg, Florian and Stutz, David and Ortiz-Jimenez, Guillermo and Kouridi, Christina and Vecerik, Mel and Hayes, Jamie and Rebuffi, Sylvestre-Alvise and Bernard, Paul and others},
  journal={arXiv preprint arXiv:2510.09263},
  year={2025}
}

@article{pan2024leveraging,
  title={Leveraging catastrophic forgetting to develop safe diffusion models against malicious finetuning},
  author={Pan, Jiadong and Gao, Hongcheng and Wu, Zongyu and Hu, Taihang and Su, Li and Huang, Qingming and Li, Liang},
  journal={Advances in Neural Information Processing Systems},
  year={2024}
}

@article{conmy2023towards,
  title={Towards automated circuit discovery for mechanistic interpretability},
  author={Conmy, Arthur and Mavor-Parker, Augustine and Lynch, Aengus and Heimersheim, Stefan and Garriga-Alonso, Adri{\`a}},
  journal={Advances in Neural Information Processing Systems},
  year={2023}
}

@inproceedings{qifine,
  title={Fine-tuning Aligned Language Models Compromises Safety, Even When Users Do Not Intend To!},
  author={Qi, Xiangyu and Zeng, Yi and Xie, Tinghao and Chen, Pin-Yu and Jia, Ruoxi and Mittal, Prateek and Henderson, Peter},
  booktitle={International Conference on Learning Representations},
  year={2024}
}

@article{thaker2025membership,
  title={Membership Inference Attacks for Unseen Classes},
  author={Thaker, Pratiksha and Kale, Neil and Wu, Zhiwei Steven and Smith, Virginia},
  journal={arXiv preprint arXiv:2506.06488},
  year={2025}
}

@article{templeton2024scaling,
       title={Scaling Monosemanticity: Extracting Interpretable Features from Claude 3 Sonnet},
       author={Templeton, Adly and Conerly, Tom and Marcus, Jonathan and Lindsey, Jack and Bricken, Trenton and Chen, Brian and Pearce, Adam and Citro, Craig and Ameisen, Emmanuel and Jones, Andy and Cunningham, Hoagy and Turner, Nicholas L and McDougall, Callum and MacDiarmid, Monte and Tamkin, Alex and Durmus, Esin and Hume, Tristan and Mosconi, Francesco
       and Freeman, C. Daniel and Sumers, Theodore R. and Rees, Edward and Batson, Joshua and Jermyn, Adam and Carter, Shan and Olah, Chris and Henighan, Tom},
       year={2024},
       journal={Transformer Circuits Thread},
       publisher={Anthropic},
       url={https://transformer-circuits.pub/2024/scaling-monosemanticity/index.html}
    }

@inproceedings{li2025detect,
  title={Detect-and-Guide: Self-regulation of Diffusion Models for Safe Text-to-Image Generation via Guideline Token Optimization},
  author={Li, Feifei and Zhang, Mi and Sun, Yiming and Yang, Min},
  booktitle={IEEE/CVF Conference on Computer Vision and Pattern Recognition},
  year={2025}
}

@misc{anthropic2024modeldiff,
author = {{Anthropic Interpretability Team}},
title = {Stage-Wise Model Diffing},
year = {2024},
url = {https://transformer-circuits.pub/2024/model-diffing/index.html},
urldate = {2025-10-02}
}

@article{minder2026overcoming,
  title={Overcoming sparsity artifacts in crosscoders to interpret chat-tuning},
  author={Minder, Julian and Dumas, Cl{\'e}ment and Juang, Caden and Chughtai, Bilal and Nanda, Neel},
  journal={Advances in Neural Information Processing Systems},
  volume={38},
  pages={106423--106474},
  year={2026}
}

@inproceedings{zhuang2023pilot,
  title={A pilot study of query-free adversarial attack against stable diffusion},
  author={Zhuang, Haomin and Zhang, Yihua and Liu, Sijia},
  booktitle={IEEE/CVF Conference on Computer Vision and Pattern Recognition},
  pages={2385--2392},
  year={2023}
}

@inproceedings{kim2024race,
  title={Race: Robust adversarial concept erasure for secure text-to-image diffusion model},
  author={Kim, Changhoon and Min, Kyle and Yang, Yezhou},
  booktitle={European Conference on Computer Vision},
  year={2024},
}

@misc{reportact,
  title = "{REPORT Act}",
  author = {{U.S. Senate}},
  howpublished = {S.474 - 118th Congress (2023-2024)},
  year = {2024},
  url = {https://www.congress.gov/bill/118th-congress/senate-bill/474}
}

@article{henderson2024safety,
  title={Safety Risks from Customizing Foundation Models via Fine-tuning},
  author={Henderson, Peter and Qi, Xiangyu and Zeng, Yi and Xie, Tinghao and Chen, Pin-Yu and Jia, Ruoxi and Mittal, Prateek},
  journal={Policy Brief. Stanford Human-Centered Artificial Intelligence},
  year={2024}
}

@inproceedings{henderson2023self,
  title={Self-destructing models: Increasing the costs of harmful dual uses of foundation models},
  author={Henderson, Peter and Mitchell, Eric and Manning, Christopher and Jurafsky, Dan and Finn, Chelsea},
  booktitle={AAAI/ACM Conference on AI, Ethics, and Society},
  year={2023}
}

@article{zhang2024defensive,
  title={Defensive unlearning with adversarial training for robust concept erasure in diffusion models},
  author={Zhang, Yimeng and Chen, Xin and Jia, Jinghan and Zhang, Yihua and Fan, Chongyu and Liu, Jiancheng and Hong, Mingyi and Ding, Ke and Liu, Sijia},
  journal={Advances in Neural Information Processing Systems},
  year={2024}
}

@misc{CompVis2022StableDiffusion,
  author       = {CompVis},
  title        = {Stable Diffusion Safety Checker},
  year         = {2022},
  howpublished = {\url{https://huggingface.co/CompVis/stable-diffusion-safety-checker}},
  note         = {Accessed: 2025-03-21}
}

@inproceedings{rando2022red,
  title={Red-teaming the stable diffusion safety filter},
  author={Rando, Javier and Paleka, Daniel and Lindner, David and Heim, Lennart and Tram{\`e}r, Florian},
  year={2022},
  booktitle = {NeurIPS 2022 ML Safety Workshop},
}

@inproceedings{rombach2022high,
  title={High-resolution image synthesis with latent diffusion models},
  author={Rombach, Robin and Blattmann, Andreas and Lorenz, Dominik and Esser, Patrick and Ommer, Bj{\"o}rn},
  booktitle={IEEE/CVF Conference on Computer Vision and Pattern Recognition},
  year={2022}
}

@inproceedings{gandikota2023erasing,
  title={Erasing concepts from diffusion models},
  author={Gandikota, Rohit and Materzynska, Joanna and Fiotto-Kaufman, Jaden and Bau, David},
  booktitle={IEEE/CVF international conference on computer vision},
  year={2023}
}

@article{yuan2025lurks,
  title={What Lurks Within? Concept Auditing for Shared Diffusion Models at Scale},
  author={Yuan, Xiaoyong and Ma, Xiaolong and Guo, Linke and Zhang, Lan},
  journal={ACM Conference on Computer and Communications Security},
  year={2025}
}

@article{guan2024deliberative,
  title={Deliberative alignment: Reasoning enables safer language models},
  author={Guan, Melody Y and Joglekar, Manas and Wallace, Eric and Jain, Saachi and Barak, Boaz and Helyar, Alec and Dias, Rachel and Vallone, Andrea and Ren, Hongyu and Wei, Jason and others},
  journal={arXiv preprint arXiv:2412.16339},
  year={2024}
}

@article{zou2023universal,
  title={Universal and transferable adversarial attacks on aligned language models},
  author={Zou, Andy and Wang, Zifan and Carlini, Nicholas and Nasr, Milad and Kolter, J Zico and Fredrikson, Matt},
  journal={arXiv preprint arXiv:2307.15043},
  year={2023}
}

@inproceedings{markov2023holistic,
  title={A holistic approach to undesired content detection in the real world},
  author={Markov, Todor and Zhang, Chong and Agarwal, Sandhini and Nekoul, Florentine Eloundou and Lee, Theodore and Adler, Steven and Jiang, Angela and Weng, Lilian},
  booktitle={AAAI Conference on Artificial Intelligence},
  year={2023}
}

@inproceedings{robey2023smoothllm,
  title={Smoothllm: Defending large language models against jailbreaking attacks},
  author={Robey, Alexander and Wong, Eric and Hassani, Hamed and Pappas, George J},
  booktitle = {Transactions on Machine Learning Research},
  year={2025}
}

@article{jain2023baseline,
  title={Baseline defenses for adversarial attacks against aligned language models},
  author={Jain, Neel and Schwarzschild, Avi and Wen, Yuxin and Somepalli, Gowthami and Kirchenbauer, John and Chiang, Ping-yeh and Goldblum, Micah and Saha, Aniruddha and Geiping, Jonas and Goldstein, Tom},
  journal={arXiv preprint arXiv:2309.00614},
  year={2023}
}

@article{wei2023jailbreak,
  title={Jailbreak and guard aligned language models with only few in-context demonstrations},
  author={Wei, Zeming and Wang, Yifei and Li, Ang and Mo, Yichuan and Wang, Yisen},
  journal={arXiv preprint arXiv:2310.06387},
  year={2023}
}

@article{inan2023llama,
  title={Llama guard: Llm-based input-output safeguard for human-ai conversations},
  author={Inan, Hakan and Upasani, Kartikeya and Chi, Jianfeng and Rungta, Rashi and Iyer, Krithika and Mao, Yuning and Tontchev, Michael and Hu, Qing and Fuller, Brian and Testuggine, Davide and others},
  journal={arXiv preprint arXiv:2312.06674},
  year={2023}
}

@inproceedings{ma2024jailbreaking,
  title={Jailbreaking prompt attack: A controllable adversarial attack against diffusion models},
  author={Ma, Jiachen and Li, Yijiang and Xiao, Zhiqing and Cao, Anda and Zhang, Jie and Ye, Chao and Zhao, Junbo},
  booktitle={Findings of the Association for Computational Linguistics},
  year={2025}
}

@article{li2025dream,
  title={DREAM: Scalable Red Teaming for Text-to-Image Generative Systems via Distribution Modeling},
  author={Li, Boheng and Wang, Junjie and Li, Yiming and Hu, Zhiyang and Qi, Leyi and Dong, Jianshuo and Wang, Run and Qiu, Han and Qin, Zhan and Zhang, Tianwei},
  journal={arXiv preprint arXiv:2507.16329},
  year={2025}
}

@article{li2024art,
  title={ART: automatic red-teaming for text-to-image models to protect benign users},
  author={Li, Guanlin and Chen, Kangjie and Zhang, Shudong and Zhang, Jie and Zhang, Tianwei},
  journal={Advances in Neural Information Processing Systems},
  year={2024}
}

@misc{grossman2025csam,
  author = {Grossman, S. and Pfefferkorn, R. and Liu, S.},
  title = {AI-Generated Child Sexual Abuse Material: Insights from Educators, Platforms, Law Enforcement, Legislators, and Victims. Version 1},
  year = {2025},
  publisher = {Stanford Digital Repository},
  url = {https://purl.stanford.edu/mn692xc5736/version/1},
  note = {Accessed: 2025-09-11}
}

@article{daras2022discovering,
  title={Discovering the hidden vocabulary of dalle-2},
  author={Daras, Giannis and Dimakis, Alexandros G},
  journal={arXiv preprint arXiv:2206.00169},
  year={2022}
}

@inproceedings{chin2023prompting4debugging,
  title={Prompting4debugging: Red-teaming text-to-image diffusion models by finding problematic prompts},
  author={Chin, Zhi-Yi and Jiang, Chieh-Ming and Huang, Ching-Chun and Chen, Pin-Yu and Chiu, Wei-Chen},
  booktitle={International Conference on Machine Learning},
  year={2024}
}

@misc{google2023red,
  author = {Google},
  title = {Google’s AI Red Team: The Ethical Hackers Making AI Safer},
  year = {2023},
  month = {July},
  day = {19},
  url = {https://blog.google/technology/safety-security/googles-ai-red-team-the-ethical-hackers-making-ai-safer},
  note = {Accessed: 2023-10-27}
}

@article{ganguli2022red,
  title={Red teaming language models to reduce harms: Methods, scaling behaviors, and lessons learned},
  author={Ganguli, Deep and Lovitt, Liane and Kernion, Jackson and Askell, Amanda and Bai, Yuntao and Kadavath, Saurav and Mann, Ben and Perez, Ethan and Schiefer, Nicholas and Ndousse, Kamal and others},
  journal={arXiv preprint arXiv:2209.07858},
  year={2022}
}

@article{tallam2025removing,
  title={Removing Watermarks with Partial Regeneration using Semantic Information},
  author={Tallam, Krti and Cava, John Kevin and Geniesse, Caleb and Erichson, N Benjamin and Mahoney, Michael W},
  journal={arXiv preprint arXiv:2505.08234},
  year={2025}
}

@article{pang2024no,
  title={No free lunch in llm watermarking: Trade-offs in watermarking design choices},
  author={Pang, Qi and Hu, Shengyuan and Zheng, Wenting and Smith, Virginia},
  journal={Advances in Neural Information Processing Systems},
  year={2024}
}

@article{wan2022comprehensive,
  title={A comprehensive survey on robust image watermarking},
  author={Wan, Wenbo and Wang, Jun and Zhang, Yunming and Li, Jing and Yu, Hui and Sun, Jiande},
  journal={Neurocomputing},
  year={2022},
  publisher={Elsevier}
}

@inproceedings{gao2024meta,
  title={Meta-unlearning on diffusion models: Preventing relearning unlearned concepts},
  author={Gao, Hongcheng and Pang, Tianyu and Du, Chao and Hu, Taihang and Deng, Zhijie and Lin, Min},
  booktitle={IEEE/CVF International Conference on Computer Vision},
  year={2025}
}

@inproceedings{fernandez2023stable,
  title={The stable signature: Rooting watermarks in latent diffusion models},
  author={Fernandez, Pierre and Couairon, Guillaume and J{\'e}gou, Herv{\'e} and Douze, Matthijs and Furon, Teddy},
  booktitle={IEEE/CVF International Conference on Computer Vision},
  year={2023}
}

@article{wen2023tree,
  title={Tree-ring watermarks: Fingerprints for diffusion images that are invisible and robust},
  author={Wen, Yuxin and Kirchenbauer, John and Geiping, Jonas and Goldstein, Tom},
  journal={Advances in Neural Information Processing Systems},
  year={2023}
}

@inproceedings{yu2021artificial,
  title={Artificial fingerprinting for generative models: Rooting deepfake attribution in training data},
  author={Yu, Ning and Skripniuk, Vladislav and Abdelnabi, Sahar and Fritz, Mario},
  booktitle={IEEE/CVF International conference on computer vision},
  year={2021}
}

@misc{comfy2025,
  author       = {{Comfy}},
  title        = {{ComfyUI | Generate video, images, 3D, audio with AI}},
  year         = {2025},
  url          = {https://www.comfy.org/},
  urldate      = {2025-09-08},
  howpublished = {\url{https://www.comfy.org/}},
  note         = {Accessed: 2025-09-08}
}

@conference{min2023silo,
  title={Silo language models: Isolating legal risk in a nonparametric datastore},
  author={Min, Sewon and Gururangan, Suchin and Wallace, Eric and Shi, Weijia and Hajishirzi, Hannaneh and Smith, Noah A and Zettlemoyer, Luke},
  booktitle={International Conference on Learning Representations},
  year={2024}
}

@article{hendrycks2023overview,
  title={An overview of catastrophic AI risks},
  author={Hendrycks, Dan and Mazeika, Mantas and Woodside, Thomas},
  journal={arXiv preprint arXiv:2306.12001},
  year={2023}
}

@article{obrien2025deep,
  title={Deep Ignorance: Filtering Pretraining Data Builds Tamper-Resistant Safeguards into Open-Weight LLMs},
  author={O'Brien, Kyle and Casper, Stephen and Anthony, Quentin and Korbak, Tomek and Kirk, Robert and Davies, Xander and Mishra, Ishan and Irving, Geoffrey and Gal, Yarin and Biderman, Stella},
  journal={arXiv preprint arXiv:2508.06601},
  year={2025}
}

@inproceedings{lu2024mace,
  title={Mace: Mass concept erasure in diffusion models},
  author={Lu, Shilin and Wang, Zilan and Li, Leyang and Liu, Yanzhu and Kong, Adams Wai-Kin},
  booktitle={IEEE/CVF Conference on Computer Vision and Pattern Recognition},
  year={2024}
}

@inproceedings{maini2025safety,
  title={Safety pretraining: Toward the next generation of safe ai},
  author={Maini, Pratyush and Goyal, Sachin and Sam, Dylan and Robey, Alex and Savani, Yash and Jiang, Yiding and Zou, Andy and Fredrikson, Matt and Lipton, Zachary C and Kolter, J Zico},
  booktitle={Advances in Neural Information Processing Systems},
  year={2025}
}

@article{okawa2023compositional,
  title={Compositional abilities emerge multiplicatively: Exploring diffusion models on a synthetic task},
  author={Okawa, Maya and Lubana, Ekdeep S and Dick, Robert and Tanaka, Hidenori},
  journal={Advances in Neural Information Processing Systems},
  year={2023}
}

@inproceedings{zhang2021can,
  title={Can subnetwork structure be the key to out-of-distribution generalization?},
  author={Zhang, Dinghuai and Ahuja, Kartik and Xu, Yilun and Wang, Yisen and Courville, Aaron},
  booktitle={International Conference on Machine Learning},
  year={2021},
  organization={PMLR}
}

@misc{googlecst,
  author = {Susan Jasper},
  title = {How we detect, remove and report child sexual abuse material},
  year = {2022},
  howpublished = {\url{https://blog.google/technology/safety-security/how-we-detect-remove-and-report-child-sexual-abuse-material/}},
  publisher = {Google}
}

@inproceedings{gutfeter2023detecting,
  title={Detecting sexually explicit content in the context of the child sexual abuse materials (CSAM): end-to-end classifiers and region-based networks},
  author={Gutfeter, Weronika and Gajewska, Joanna and Pacut, Andrzej},
  booktitle={Joint European Conference on Machine Learning and Knowledge Discovery in Databases},
  year={2023},
}

@article{lee2020detecting,
  title={Detecting child sexual abuse material: A comprehensive survey},
  author={Lee, Hee-Eun and Ermakova, Tatiana and Ververis, Vasilis and Fabian, Benjamin},
  journal={Forensic Science International: Digital Investigation},
  volume={34},
  pages={301022},
  year={2020},
  publisher={Elsevier}
}

@article{bommasani2021opportunities,
  title={On the opportunities and risks of foundation models},
  author={Bommasani, Rishi and others},
  journal={arXiv preprint arXiv:2108.07258},
  year={2021}
}

@article{magid2024you,
  title={Is What You Ask For What You Get? Investigating Concept Associations in Text-to-Image Models},
  author={Magid, Salma S Abdel and Pan, Weiwei and Warchol, Simon and Guo, Grace and Kim, Junsik and Rahman, Mahia and Pfister, Hanspeter},
  journal={Transactions on Machine Learning Research},
  year={2025}
}

@article{thiel2023generative,
  title={Generative ML and CSAM: Implications and mitigations},
  author={Thiel, David and Stroebel, Melissa and Portnoff, Rebecca},
  journal={Stanford Digital Repository},
  year={2023}
}

@misc{thornsafetybydesign,
title={Safety by Design for Generative AI: Preventing Child Sexual Abuse},
howpublished={\url{https://info.thorn.org/hubfs/thorn-safety-by-design-for-generative-AI.pdf}},
author={Thorn and {All Tech is Human}},
year=2024
}

@misc{2025report,
  title={The Deepfake Dilemma: New challenges protecting students, confidentiality},
  author={Patricia Davis},
  year={2025},
  howpublished={\url{https://www.missingkids.org/blog/2025/the-deepfake-dilemma-new-challenges-protecting-students-confidentiality}},
  publisher={National Center for Missing & Exploited Children (NCMEC)}
}

@misc{thorn24,
  title={Youth Perspectives on Online
Safety},
  author={Thorn},
  year={2024},
  howpublished={\url{https://info.thorn.org/hubfs/Research/Thorn_23_YouthMonitoring_Report.pdf}},
  publisher={Thorn}
}

@inproceedings{
hazra2025position,
title={Position: {AI} Safety should prioritize the Future of Work},
author={Sanchaita Hazra and Bodhisattwa Prasad Majumder and Tuhin Chakrabarty},
booktitle={International Conference on Machine Learning, Position Paper Track},
year={2025},
}

@inproceedings{peppin2025reality,
  title={The reality of ai and biorisk},
  author={Peppin, Aidan and Reuel, Anka and Casper, Stephen and Jones, Elliot and Strait, Andrew and Anwar, Usman and Agrawal, Anurag and Kapoor, Sayash and Koyejo, Sanmi and Pellat, Marie and others},
  booktitle={ACM Conference on Fairness, Accountability, and Transparency},
  year={2025}
}

@article{musser2023cost,
  title={A cost analysis of generative language models and influence operations},
  author={Musser, Micah},
  journal={arXiv preprint arXiv:2308.03740},
  year={2023}
}

@article{hazell2023spear,
  title={Spear phishing with large language models},
  author={Hazell, Julian},
  journal={arXiv preprint arXiv:2305.06972},
  year={2023}
}

@article{birhane2023into,
  title={Into the laion’s den: Investigating hate in multimodal datasets},
  author={Birhane, Abeba and Prabhu, Vinay and Han, Sanghyun and Boddeti, Vishnu and Luccioni, Sasha},
  journal={Advances in Neural Information Processing Systems, Track on Datasets and Benchmarks},
  year={2023}
}

@inproceedings{kapoor2024societal,
  title={On the societal impact of open foundation models},
  author={Kapoor, Sayash and Bommasani, Rishi and Klyman, Kevin and Longpre, Shayne and Ramaswami, Ashwin and Cihon, Peter and Hopkins, Aspen and Bankston, Kevin and Biderman, Stella and Bogen, Miranda and others},
  booktitle={International Conference on Machine Learning, Position Paper Track},
  year={2024}
}

@misc{internet2023ai,
  title={How AI is being abused to create child sexual abuse imagery},
  author={IWF},
  year={2023},
  howpublished={\url{https://www.iwf.org.uk/about-us/why-we-exist/our-research/how-ai-is-being-abused-to-create-child-sexual-abuse-imagery/}},
  publisher={Internet Watch Foundation}
}

@misc{iwf2024,
  title={What has changed in the AI CSAM landscape?},
  author={IWF},
  year={2024},
  howpublished={\url{https://www.iwf.org.uk/media/nadlcb1z/iwf-ai-csam-report_update-public-jul24v13.pdf}},
  publisher={Internet Watch Foundation}
}

@misc{iwfblog,
  title={Full feature-length AI films of child sexual abuse will be ‘inevitable’ as synthetic videos make ‘huge leaps’ in sophistication in a year},
  author={IWF},
  year={2025},
  howpublished={\url{https://www.iwf.org.uk/news-media/news/full-feature-length-ai-films-of-child-sexual-abuse-will-be-inevitable-as-synthetic-videos-make-huge-leaps-in-sophistication-in-a-year/}},
  publisher={Internet Watch Foundation}
}

@misc{paltieli,
  title={How Predators Are Abusing Generative AI},
  author={Paltieli, Guy and Freud, Gideon},
  url = {https://www.
activefence.com/blog/predators-abusing-generative-ai},
  year={2023},
  publisher={ActiveFence}
}

@article{thiel2023identifying,
  title={Identifying and eliminating csam in generative ml training data and models},
  author={Thiel, David},
  journal={Stanford Internet Observatory, Cyber Policy Center},
  year={2023}
}

@misc{USC18_possession,
    title = {Certain activities relating to material involving the sexual exploitation of minors},
    author = {{US Congress}},
    howpublished = {18 U.S.C. \S 2252},
    year = {2011},
    url = {https://uscode.house.gov/view.xhtml?req=granuleid:USC-prelim-title18-section2252&num=0&edition=prelim}
}

@misc{USC18_protect,
    title = {Prosecutorial Remedies and Other Tools to end the Exploitation of Children Today Act},
    author = {{US Congress}},
    howpublished = {S. 151},
    year = {2003},
    url = {https://www.congress.gov/bill/108th-congress/senate-bill/151}
}

@misc{thornsafetybydesignblog,
  title={Thorn and All Tech Is Human Forge Generative AI Principles with AI Leaders to Enact Strong Child Safety Commitments},
  author={Thorn},
  howpublished="\url{https://www.thorn.org/blog/generative-ai-principles/}",
  year={2024},
  publisher={Thorn}
}

@misc{eSafetytransparencyreport0825,
  title={A baseline for online safety transparency},
  author={eSafety},
howpublished="\url{https://www.esafety.gov.au/sites/default/files/2025-08/BOSE-full-report-CSEA-sexual-extortion-periodic-notices-August2025.pdf?v=1757703246816}",
  year={2025},
  publisher={eSafety}
}

@article{spence23,
  title={Content Moderators’ Strategies for Coping with the Stress of Moderating Content Online},
  author={Ruth Spence and Amy Harrison and Paula Bradbury and Paul Bleakley and Elena Martellozzo and and Jeffrey DeMarco},
  journal={Journal of Online Trust and Safety},
  volume={1},
  pages={5},
  year={2023}
}

@misc{TakeItDownAct,
    title = {TAKE IT DOWN Act},
    author = {{US Congress}},
    howpublished = {S.146},
    year = {2025},
    url = {https://www.congress.gov/bill/119th-congress/senate-bill/146}
}

@article{meta,
title={Meta files lawsuit against developer of CrushAI ‘nudify’ app},
author={Jonathan Vanian},
journal={CNBC},
year={2025},
url={https://www.cnbc.com/2025/06/12/meta-files-lawsuit-against-developer-of-crushai-nudify-app.html}}

@misc{fbiPSA,
  title={Criminals Use Generative Artificial Intelligence to Facilitate Financial Fraud},
  author={FBI},
  year={2024},
  howpublished={\url{https://www.ic3.gov/PSA/2024/PSA241203}},
  publisher={Federal Bureau of Investigations (FBI)}
}

@misc{graphika25,
  title={Character Flaws. School Shooters, Anorexia Coaches, and Sexualized Minors: A Look at Harmful Character Chatbots and the Communities That Build Them},
  author={Cristina López G., Daniel Siegel, Erin McAweeney},
  year={2025},
  howpublished={\url{https://public-assets.graphika.com/reports/graphika-report-character-flaws.pdf}},
  publisher={Graphika}
}

@misc{horwitz2025,
	title = {Meta’s ‘Digital Companions’ Will Talk Sex With Users—Even Children},
	howpublished = {\url{https://www.wsj.com/tech/ai/meta-ai-chatbots-sex-a25311bf}},
	journal = {Wall Street Journal},
	author = {Horwitz, Jeff},
	year = {2025},
}

@article{wolbers2025,
author = {Wolbers, Heather and Cubitt, Timothy and Napier, Sarah and Cahill, Michael John and Nicholas, Mariesa and Burton, Melanie and Giunta, Katherine},
title = {Sexual extortion of Australian adolescents: Results from a national survey},
year = {2025},publisher = {Australian Institute of Criminology},
journal = {Trends and Issues in Crime and Criminal Justice},
}

@misc{thorn2025sextortion,
	title = {Sexual Extortion \&
Young People: Navigating Threats in Digital Environments},
	howpublished = {\url{https://info.thorn.org/hubfs/Research/Thorn_SexualExtortionandYoungPeople_June2025.pdf}},
	author = {Thorn},
	year = {2025},
}

@misc{hf_video_generation_leaderboard,
  author       = {{Artificial Analysis}},
  title        = {Video Generation Arena Leaderboard},
  year         = {2025},
  howpublished = {HuggingFace Spaces},
  url          ={https://huggingface.co/spaces/ArtificialAnalysis/Video-Generation-Arena-Leaderboard},
  note         = {Accessed January 2026}
}

@misc{hf_text_to_image_leaderboard,
  author       = {{Artificial Analysis}},
  title        = {Text-to-Image Leaderboard},
  year         = {2025},
  howpublished = {HuggingFace Spaces},
  url          ={https://huggingface.co/spaces/ArtificialAnalysis/Text-to-Image-Leaderboard},
  note         = {Accessed January 2026}
}

\newpage
\appendix

\section{Glossary}
\label{app:glossary}

\begin{table}[h!]
  \centering
  \begin{tabular}{>{\raggedleft}p{0.24\columnwidth} p{0.65\columnwidth}}
    \toprule
            \textbf{CSEA} & Child sexual abuse and exploitation \\
            \textbf{AIG-CSAM }& AI-generated child sexual abuse material \\
            \textbf{Reporting hotlines} & National Center for Missing \& Exploited Children (NCMEC), Internet Watch Foundation (IWF) \\
            \textbf{AI Developer} & The entity responsible for designing, training, and testing an AI model or system. \\
            \textbf{AI Deployer} & The entity that puts an AI system into service and controls its operation, often by integrating it into a product or platform. \\
            \textbf{AI Provider} & Any entity that makes an AI system or model available for use. \\
    \bottomrule
 \end{tabular}
 \caption{Key definitions used throughout this paper.}
 \label{tab:terms}
\end{table}

\begin{table*}[th!]
    \centering
    \begin{tabular}{r  ccccccccccccccc}
        \toprule
        Open Problem \# & \textbf{1} & \textbf{2} & \textbf{3} & \textbf{4} & \textbf{5} & \textbf{6} & \textbf{7} & \textbf{8} & \textbf{9} & \textbf{10} & \textbf{11} & \textbf{12} & \textbf{13} & \textbf{14} & \textbf{15} \\
        \midrule
        Data restrictions & \cmark & \cmark & \cmark & \xmark & \cmark & \cmark & \xmark & \xmark & \cmark & \cmark & \xmark & \cmark & \xmark & \cmark & \xmark \\
        Evaluation restrictions & \cmark & \cmark & \cmark & \xmark & \cmark & \cmark & \cmark & \cmark & \xmark & \xmark & \cmark & \xmark & \cmark & \xmark & \cmark \\
        Adversarial environment & \xmark & \cmark & \cmark & \cmark & \cmark & \cmark & \cmark & \cmark & \cmark & \cmark & \cmark & \cmark & \cmark & \cmark & \cmark \\
        Wellness implications & \xmark & \xmark & \xmark & \cmark & \xmark & \cmark & \cmark & \cmark & \xmark & \cmark & \xmark & \xmark & \xmark & \cmark & \cmark \\
         \bottomrule
    \end{tabular}
    \caption{Key challenges and broad restrictions across the proposed open problems in preventing CSAM generation.}
    \label{tab:my_label}
\end{table*}

\section{Additional Related Work}
\label{sec:relatedwork}

Extending AI safety to AIG-CSAM prevention naturally builds on diverse prior work that mitigates AI risks. Throughout this work, we highlight many relevant papers in these domains, e.g., on topics such as on red teaming \cite{ganguli2022red}, model fingerprinting~\cite{guan2022you}, and tamper resistance \cite{henderson2023self}. In addition, several papers (1) outline broader open problems in AI safety, or (2) propose child-safety specific mitigations. We highlight these areas of related work here.

\paragraph{Open Problems in AI Safety.} \citet{casper2025open} outline open problems in derisking open-source models, several of which directly address child safety, such as reliable tamper resistance (Open Problem 3: Resilience to harmful fine-tuning). Similarly, \citet{kamachee2025video} demonstrate problems with the open-source model landscape that enable video deepfake propagation, including video CSAM. Specific to Open Problem 1 (Partial data cleaning), \citet{cretu2025evaluating} find that partial data filtering is insufficient for CSAM prevention and deconstruct open subproblems, such as accurate child detection.

\paragraph{Child Safety Mitigations.} Some existing works specifically address open problems outlined in this paper for preventing AIG-CSAM. For example, \citet{thiel2023identifying} propose annotation pipelines to improve data cleaning (Open Problem 1) and apply them to Stable Diffusion. Work from \citet{gutfeter2023detecting} develops an end-to-end CSAM classifier and proposes an evaluation methodology incorporating proxy data and controlled access to CSAM (Open Problem 6). There remains a significant need for further work on these problems.

\paragraph{Comparison to Other Restricted Domains.}

Child sexual abuse material (CSAM) is unique in that it is the only media in the United States that is illegal to create, possess, or distribute \cite{citizens2026doj}. This is also broadly true under global regulatory regimes \cite{csam2023icmec}. While there are other domains with data that may be similarly traumatic to view, CSAM is unique in its illegal status. Government-sensitive material bears some similarity in that its distribution is limited \cite{natsec1995potus}, but CSAM access is not tied to security clearance levels the way other government-sensitive data is. Medical records under HIPAA also bear similarity in that data is restricted \cite{hipaa1996congress}, but they are not subject to evaluation, generation, and wellness constraints that CSAM falls under.

 \section{Additional Open Problems}
 \label{app:additional_openproblems}

\subsection{Developing Safe Models by Design}
\label{app:development}
 \begin{pboxchipped}
{\textbf{Open Problem A4: Minimizing human exposure}}
{\chipAdv\ \chipWell}
{How does exposure to AIG-CSAM affect red teamers? How can human exposure be minimized?}
\end{pboxchipped}
\label{sec:develop:rt-human-exposure}

The emotional and mental toll of CSAM exposure is well documented across moderation \cite{spence2023psychological}, law enforcement work \cite{rimer2025once}, and data labeling \cite{guardian2023aitoll}, with additional wellness implications for red teamers \cite{zhang2024human}. These harms demonstrate the clear need for red teaming solutions that minimize human exposure to CSAM.

\textit{Existing work.} Several frameworks for automated red teaming text-to-image models have been proposed \cite{chin2023prompting4debugging, li2024art, li2025dream}, and existing industry content review tools incorporate wellness features like image blurring \cite{das2020fast}. 

\textit{Limitations.} Text-to-image models are particularly vulnerable to out-of-distribution prompts, requiring robust human red teaming \cite{daras2022discovering, rando2022red}. Offenders quickly discover new jailbreaking mechanisms, making it challenging to ensure testing remains relevant. Moreover, evaluating wellness features for red-teamers requires sociotechnical studies with industry and red teaming service providers, who may not have incentive to participate in such studies. 

\begin{pboxchipped}
{\textbf{Open Problem A5: Watermark robustness}}
{\chipData\ \chipEval\ \chipAdv}
{How can text-to-image watermarks be made robust to removal, spoofing, and partial edits?}
\end{pboxchipped}
\label{sec:develop:watermarks}

In the CSAM space, offenders actively combat safety interventions, including stripping metadata and other identifying factors in the image (e.g. watermarks). 
Researchers have explored robust watermarking techniques that are less susceptible to spoofing or removal \cite{wan2022comprehensive, gowal2025synthid}. Methods like Stable Signature incorporate the watermark directly in the model weights during pretraining \citet{fernandez2023stable}.




\emph{Limitations.} Many schemes remain vulnerable to simple attacks \cite{pang2024no}. Watermarks can be removed by partially modifying an image \cite{tallam2025removing} or 
editing code in image generation scripts \cite{rombach2022high}. Watermarking techniques that are more robust to fine-tuning and erasure tend to be more vulnerable to being stolen and spoofed on other images \cite{pang2024no}. Robust solutions that include a history of partial edits are also lacking.

\subsubsection{Experiment Details: Concept Fusion}
\label{sec:experiment-details}

To demonstrate concept fusion, we trained conditional flow matching models 
on 128x128 CelebA images. Each model comprised a 295M parameter UNet and was trained for 700 epochs. Given two attribute classes $A, B$ (e.g., blonde hair, eyeglasses), models were trained on images from $A \setminus B \cup B \setminus A$---for instance, blondes \textit{without} eyeglasses and non-blondes \textit{with} eyeglasses (see Section~\ref{sec:conceptfusion} for examples). We train on 4K images of a certain hair color (blonde, black) without eyeglasses and a varying number of images (64-4K) without that hair color but with eyeglasses.

\begin{figure}[!ht]
    \centering
    \includegraphics[width=\linewidth]{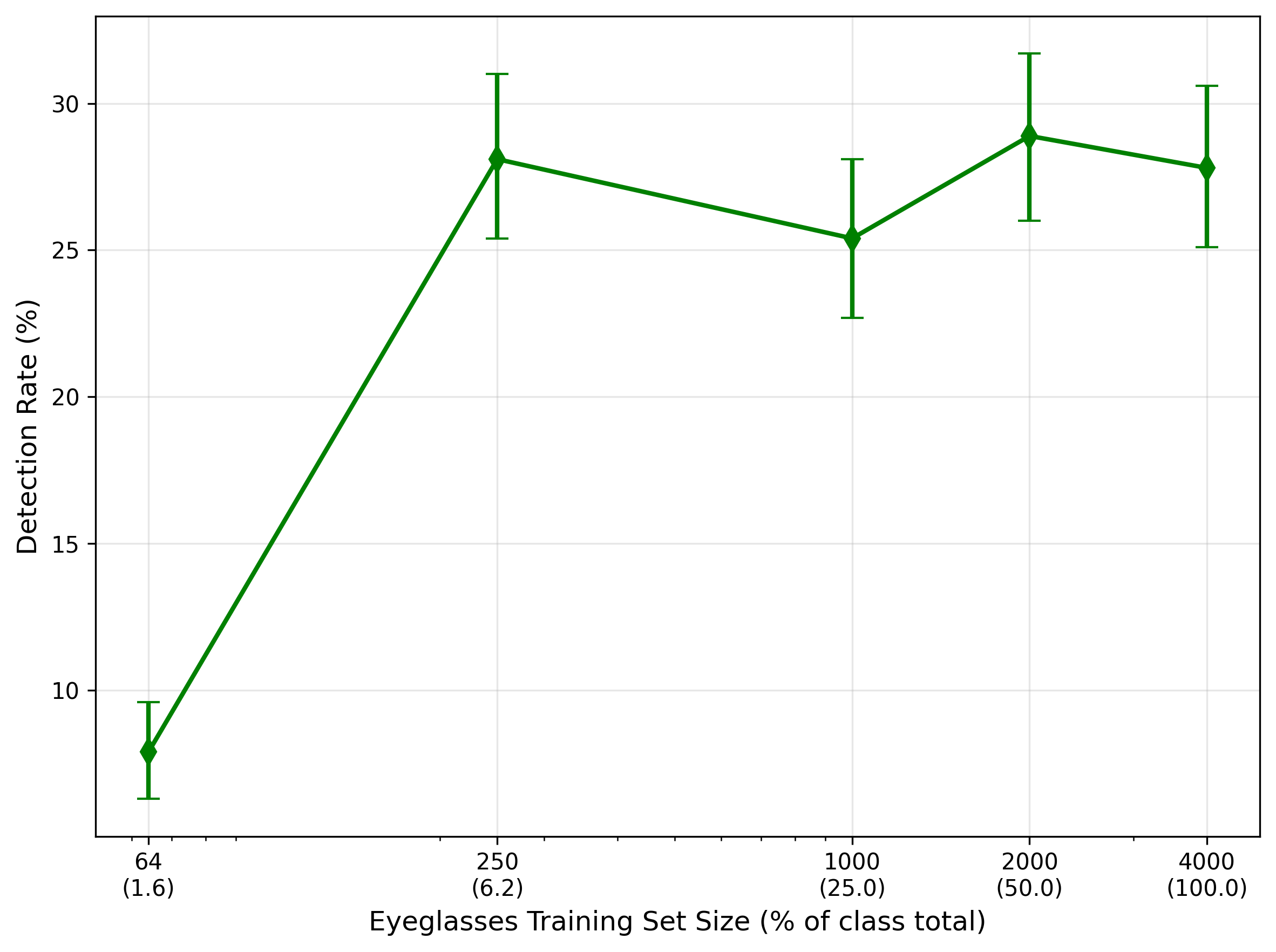}
    \caption{We test composition of the black hair and eyeglasses concepts, varying the number of eyeglasses samples from 64 to 4K. When increasing from 64 to 250 samples, capability to produce compositional images triples from 9\% to 27\%.}
    \label{fig:composition_samples_2}
\end{figure}

We measured the models' propensity to generate \textit{compositional} images from $A \cap B$ (blonde hair and eyeglasses in Figure~\ref{fig:composition_samples}, black hair and eyeglasses in Figure~\ref{fig:composition_samples_2}) while varying dataset composition. In Figure~\ref{fig:composition_samples}, we test two generation methods of (1) unconditional generation or (2) conditioning on an average class vector. We then take the maximum detection rate over these two methods. We observe a sharp threshold in the number of samples required for composition: with 750 eyeglasses samples, composition likelihood was ${\sim}0.1\%$; with 1000 samples, this increased $15\times$ to $1.5\%$.

Varying the sampling strategy can further increase this ratio. In Figure~\ref{fig:composition_samples_2}, we evaluate a sequential conditioning approach: instead of denoising unconditionally for $T$ steps, we first denoise for $n < T$ steps conditioned on class $A$, then for $T - n$ steps conditioned on class $B$. While a fixed conditioning (unconditional or average) measures propensity, this strategy measures capability. Here, capability proves substantially stronger---the model generates images from $A \cap B$ over $25\%$ of the time with just 250 eyeglasses samples ($6.2\%$ of training data).

Our findings align with \citet{okawa2023compositional}, who also observe a sharp sample threshold for concept composition. This suggests partial data cleaning may be effective, though the bar is high: in our setting, just 64 images of a ``cleaned concept'' still enable composition ${\sim}10\%$ of the time. Future work should explore how this threshold scales with larger diffusion models. Concurrent work from \citet{cretu2025evaluating} more directly evaluates model capability to compose children with other concepts and highlights additional challenges such as child detection. We note both directions as ways to pursue child-safety research on the open problems posed here, at different levels of abstraction.

\subsection{Deployment Safeguards}
\label{app:deployment}
\begin{pboxchipped}
{\textbf{Open Problem B3: Standardized safety assessments}}
{\chipEval\ \chipAdv\ \chipWell}
{How can we standardize assessments for AIG-CSAM capabilities?}
\end{pboxchipped}

Standardized safety assessments allow for consistent and transparent model evaluation. For child safety in particular, building confidence and trust in evaluations requires assurance that the assessment is robust, and not unduly influenced by other incentives, e.g., product deadlines.

\emph{Existing work.} External red teaming and benchmarking are standard for assessing generative models. External domain expertise can help discover novel issues \cite{ahmad2025openai}. Benchmarking supports scalable reproducibility, allowing multiple models from different developers to undergo the same evaluation \cite{vidgen2024introducing}.

\emph{Limitations.} Safety benchmarking assessments may correlate with model capabilities rather than actual safety \cite{ren2024safetywashing}. Other studies highlight fundamental gaps in AI safety assessments, particularly for non-text modalities \cite{rauh2024gaps}. Given the adversarial nature of CSAM offenders and the wellness and legal barriers, static benchmarks quickly becomes outdated as offenders develop new strategies for generating AIG-CSAM.

\begin{pboxchipped}
{\textbf{Open Problem B4: Model transparency}}
{\chipData\ \chipAdv}
{How can we use model cards to encourage transparency without inadvertently enabling offenders?}
\end{pboxchipped}

Model cards are the industry standard for disclosing information about models. For AIG-CSAM, documenting child safety interventions creates a natural pause point for developers to assess safeguards.

\emph{Existing work.} Model cards are intended to provide fair assessment on a variety of human critical factors, e.g., bias \cite{mitchell2019model}. 
Model card format can influence how interpretable the information is to non-technical audiences \cite{crisan2022interactive}, enabling ethical decision-making for laypersons as well.  Even the act of filling out model cards can elicit further ethical consideration from participating developers \cite{nunes2022using}.

\emph{Limitations.} Disclosing safety interventions is only effective if deployment is actually contingent on implementing them; currently, models without CSAM safeguards can still be released. While transparency enables good-faith actors to identify well-safeguarded models, it also enables offenders to discover vulnerable ones.

 

\subsubsection{Proof-of-Concept: Model Cards}

As a second proof-of-concept experiment, we manually audit public model and system cards for 14 widely used image and video generators (Table~\ref{tab:model-card-coverage}), asking whether they document child-safety relevant risks. These 14 models were chosen based on their presence in popular video generation \cite{hf_video_generation_leaderboard} and image generation leader boards \cite{hf_text_to_image_leaderboard}. This analysis is limited to publicly available documentation and does not make assessments about any deployed safeguards, absence of documentation should not be interpreted as absence of safety measures. We mark whether a public card exists and whether it reports any evaluation targeted at children, explicit sexual content in general, and CSAM.

\begin{table}[!ht]
    \centering
    \begin{tabularx}{\linewidth}{lccccc}
        \toprule
        Model & \shortstack{Card\\Exists} & Child & Explicit & CSAM \\
        \midrule
        Sora         & \cmark & \cmark & \cmark & \cmark \\
        Sora 2       & \cmark & \cmark & \cmark & \cmark \\
        DALLE-3      & \cmark & \xmark & \cmark & \xmark \\
        Imagen 4     & \cmark & \xmark & \xmark & \xmark \\
        Veo 3.1      & \cmark & \xmark & \xmark & \xmark \\
        FLUX.1         & \cmark & \xmark & \xmark & \xmark 
        \\
        DeepSeek Janus & \cmark & \xmark & \xmark & \xmark 
        \\
        SDXL         & \cmark & \xmark & \xmark & \xmark 
        \\
        Z-Image & \cmark & \xmark & \xmark & \xmark 
        \\
        RunwayML Gen 4.5 & \xmark & -- & -- & -- \\
        Grok Imagine & \xmark & -- & -- & -- \\
        Midjourney   & \xmark & -- & -- & -- \\
        Pika         & \xmark & -- & -- & -- \\
        Luma         & \xmark & -- & -- & -- \\
        \bottomrule
    \end{tabularx}
    \caption{Presence of child safety related disclosures in public documents for selected image and video generation models. This analysis examines disclosures regarding the categories of children, explicit content, and CSAM and is limited to publicly available documentation; the absence of documentation does not necessarily imply the absence of safeguards.}
    \vspace{-2em}
        \label{tab:model-card-coverage}
\end{table}

Among the systems with model or system cards, \textit{only} OpenAI’s Sora models include dedicated sections that explicitly discuss child safety and CSAM. Several of the most widely deployed systems (e.g., Midjourney, Grok Imagine, Pika, Luma) have no public model card at all, despite their scale and popularity.

\begin{pboxchipped}
{\textbf{Open Problem B5: Hobbyist ecosystem}}
{\chipData\ \chipAdv\ \chipWell}
{How can we encourage adoption of safety best practices across the diverse ML hobbyist ecosystem?}
\end{pboxchipped}

AI safety efforts typically target industry providers rather than hobbyists who build on foundation models \cite{Thorn_PAI_CaseStudy_2024}. This leaves a gap: downstream actors may lack the resources, incentives, or oversight to maintain the CSAM safeguards built into the original model.

\emph{Existing work.} AI developers broadly recognize ethical dilemmas but often lack resources and training to navigate them \cite{griffin2025ethical}. Education-based prevention of child sexual abuse is well-studied and established as a standard approach \cite{wurtele2010partnering, patterson2022systematic}. Research on hobbyist developers highlights intellectual stimulation as a primary motivation \cite{koch2014joining}.

\emph{Limitations.} Education efforts to promote developer awareness and CSAM prevention practices remain understudied. Researching those online communities carries risks of harassment and doxing \cite{doerfler2021m}, compounded by the high-stakes nature of child sexual abuse—a topic that often prompts defensiveness and avoidance when developers confront unintended consequences of their models.

\subsection{Safe Model Maintenance}
\label{app:maintenance}
\begin{pboxchipped}
{\textbf{Open Problem C5: Assessing safeguards}}
{\chipEval\ \chipAdv \chipWell}
{How do third-party auditors and users effectively assess the efficacy of implemented safeguards?}
\end{pboxchipped}

Even where safeguards have been implemented or reported as implemented, assessing their effectiveness---individually and within the broader system context---is necessary for building trust and transparency.

\emph{Limitations.} As noted earlier, most AI safety assessments focus on individual models through red teaming or benchmarks. Mechanisms for assessing complex AI systems are lacking. Sociotechnical assessments that account for how users actually engage with platforms, including offender behavior, platform-specific risks, and cross-platform dynamics, remain uncommon.

Companies may lack incentive to provide access necessary for these studies, which are important for building shared understanding and trust. Some metrics require cross-platform measurement, such as tracing circulating AIG-CSAM from downstream tools back to source models as a proxy for safeguard robustness.

\section{Policy Tradeoffs}
\label{app:policytradeoffs}
Enacting the policy recommendations highlighted in this work may require difficult decisions and challenging tradeoffs. For example: establishing secure pathways for vetted institutions to evaluate AIG-CSAM capabilities could improve accountability and benchmarking, but would also require carefully scoped governance structures to prevent misuse, leakage, or other harms associated to exposure to highly sensitive material. When considering regulatory tools for transparency: mandatory disclosures on CSAM filtering, child-safety risk assessments, and platform prevention strategies all require clear standards for compliance auditing. In the absence of these, tradeoffs between flexibility and enforceability may result in company disclosures that are overly vague and lack meaningful detail. Even with such standards in place, challenges emerge around establishing the right level of specificity within these standards. Narrow requirements risk becoming quickly outdated as AI systems evolve, while overly broad requirements could create inconsistent enforcement. 

These examples do not reflect the full scope of tradeoffs and decisions that may need to occur when enacting policy solutions in this space. Even so: we emphasize that, while these decisions are important and challenging, \textit{not} enacting policy solutions is just as much a decision. The outcome of that decision, when it comes to AIG-CSAM, is tangibly apparent: more victims and more harm. This outcome is unacceptable.

\end{document}